\documentclass[12pt,final]{iopart}
\usepackage{graphicx}
\usepackage[section]{placeins}
\usepackage[]{caption}

\begin{document}

\title{Mechanisms of Coulomb dissociation  processes}

\author{Zolt\'an Seres \footnote{For the MoNA collaboration:  
ELTE (Budapest), NSCL(East Lansing), Atomki (Debrecen), Rykkyo (Tokyo), 
and Wigner Research Centre  for Physics (Budapest), 
in experiment NSCL \#03038.}}

\address{Wigner Research Centre for Physics, Hungarian Academy of Sciences, Hungary}

\ead{seres.zoltan@wigner.mta.hu}
\date{30 October, 2017}

\begin{abstract}
The Coulomb dissociation is studied of the $^8$Li nuclei on Pb target 
at energies  40.3 and 69.5~MeV/nucleon, in the experiment NSCL \#03038.  
The $^{6,7}$Li, $^{4,6}$He, and $^2$H fragments were identified. The resonance 
decay and the direct breakup reactions were observed. The data give experimental
evidence that the Coulomb dissociation is a two step process. The projectile 
in the approaching phase is braked down and the valence neutron gets forced oscillation. 
The increasing Coulomb force holds the projectile and brings through the 
excited states up to the closest approach point. There released, the projectile
is trapped into a primary, highly excited state in the continuum. The
lifetime of the primary excited state depends on the multipolarity of the 
deformed projectile. At intermediate energy the collision
is a sudden reaction, the valence neutron may stay ---  during the impact --- in the 
forward or backward hemisphere of the projectile nucleus, and the orbit of the
valence neutron gets prolate   or oblate, dipole or multipole deformation 
and the nucleus gets single-particle or collective excitation. The primary excited 
states, developing will decay prompt or delayed into the same reaction channel, 
resulting in the two decay mechanisms. 
%
%
%
\end{abstract}


\section{Introduction}
\label{intr}

In large impact parameter collisions, when the closest approach 
distance is larger than the sum of the two radii, the Coulomb 
interaction may induce nuclear reactions. In the experiment NSCL 
\#03038 the Pb($^8$Li,$^7$Li+n)Pb Coulomb breakup reaction was investigated 
because of its astrophysical interest  \cite{Ho,Iz1} and to study the 
mechanism of the  reaction. The intermediate and high energy light ions 
in the strong electric field of a high Z target nucleus may suffer 
Coulomb dissociation. The Coulomb breakup gives sharp peaks  above 
the particle threshold with large transition strength of long lived, 
low energy unbound resonance states, the soft dipole resonances \cite{Ha,Ga}. 
However, in other experiments there were observed fragments faster than 
the beam  \cite{Ie,Sa}. Supposedly those decayed in the Coulomb field of 
the target nucleus and after the decay the fragments were post-breakup 
accelerated with Z/A changed. It requests that the lifetime of the 
excited state   must be many orders of magnitude less than the lifetime  
calculated by the uncertainty  principle from the width of the known 
unbound resonance levels. This paper analyzes the data of the above 
experiment in order to investigate the reaction mechanisms of the 
Coulomb dissociation. 

The post-acceleration was  not found in every experiment.
In the  Coulomb dissociation of 41~MeV/nucleon $^{11}$Be there could not be
verified the post-acceleration in the longitudinal velocity of the fragments
\cite{An,BU}. But  K. Riisager {\it et al.} \cite{Ri} found that the neutron velocity  
was shifted down, emitted from the decelerated projectile in the Coulomb dissociation 
of 29~MeV/nucleon $^{11}$Li on gold target.   D. Sacket {\it et al.}
\cite{Sk,Ie} found that the velocity difference of the fragments and neutrons was 
definitely larger than zero   ($ \Delta v = v_9- v_{2n}=0.0090 \pm 0.0003$~c) 
in the Coulomb breakup of 28~MeV/nucleon $^{11}$Li on Pb target. They explained
it with post-breakup acceleration of the fragment.

A lot of theoretic effort was invested to study the post-acceleration of the fragments.
G. Baur {\it et al.} shew that the longitudinal post-breakup acceleration of the fragment 
can be estimated in a classical model, however, it was absent in quantal model based
on Glauber theory in sudden approximation \cite{BaBe}.  The higher order electromagnetic 
interaction may play decisive role in the post-breakup acceleration \cite{BeBa}. G. F. Bertsch 
and C. A. Bertulani \cite{BB}  calculated the longitudinal momentum shift of the fragment 
of the $^{11}$Li$\rightarrow^9$Li+2n Coulomb dissociation at 28~MeV/nucleon energy. 
They solved the time-dependent Schr\"odinger equation in one dimension, and got 
post-acceleration of the fragment in the order of magnitude of the measured value. 
They compared the quantum prediction with the classical model of instantaneous 
decay  where the neutron is fixed to the projectile up to the closest approach
point and there decays. The quantum and the instantaneous models give similar 
post-acceleration values in the 20--50~fm impact parameter range. Sagawa {\it et al.} 
\cite{Sa} studied the multipole excitation of the valence nuclei. They concluded that  
the extremely large transition strength, the sharp peak above the particle threshold  
is non  collective, single-particle excitation.  T. Kido {\it et al.} \cite{Ki}  solved 
numerically the time dependent Schr\"odinger equation in the external Coulomb field 
of the target nucleus. They took a single neutron and the core, expanding the external 
field into multipoles. From the time evolution the post-breakup acceleration, 
the relative longitudinal momentum shift was under estimated by a factor of 2. 

Some statements in the upper papers serve relevance for the conclusions derived from the
 data of the present experiment as follows.  At intermediate and high energy the Coulomb 
disintegration is a {\it sudden process}, for the beam velocity is much larger than 
the intrinsic motion of the valence neutron \cite{An,BaBe,BB}. The dissociation on heavy 
target, at large impact parameter  is {\it dominated by the Coulomb excitation}, the 
contribution of the Coulomb-nuclear interference and nuclear interaction 
is small at impact parameters in the order of 20~fm \cite{An,Be}.
The projectile is {\it excited direct to the continuum}. The excited states concentrate 
in a {\it narrow region around the giant resonances} \cite{Be}. The resonances {\it decay 
mostly by particle}, or are attenuated by $\gamma $-ray emission (at a few tenth of MeV/nucleon 
energy the  $\gamma $-ray multiplicity is about 10 \cite{Mo}). 
The {\it time delay} of the particle emission can play an important 
role in the Coulomb dissociation process \cite{BB}. The {\it instantaneous decay model} 
gives realistic result on the contrary the raw classical approximation \cite{BB}.

The experimental setup is described in Sec.~2. The acceptance of 
the fragment detector and the magnetic field cut small phase space 
cells separating several fragments of different Z and A, momenta and
scattering angle, {\it i.e.} several reaction channels.
The fragment identification is described in Sec.~3.  
At 40~MeV/nucleon the $^7$Li, $^6$Li, $^6$He, and $^4$He, at 70~MeV/nucleon 
the $^7$Li, $^6$Li, $^4$He, and $^2$H fragments can be separated.
The yield of the reaction channels is discussed inc Sec.~4.
The two reaction mechanisms are presented and verified in Sec.~5.
The decays of the long-lived unbound resonance states are discussed
in Sec.~6. Low-lying resonance states are  identified  
of the $^8$Li, $^7$Li, and $^6$Li nuclei. The direct breakup 
reaction is analyzed in Sec.~7. The neutron and fragment velocity spectra  
are compared and their correlation is investigated. It was found that 
the neutron velocity of the  resonance and the direct breakup reaction
has identical structure, the reaction products are from the same 
final transitions. The correlation of the velocities verifies the 
identification of the fragment charge. 

An explanation of the observed phenomena is proposed in Sec.~8.
It is Supposed that the Coulomb dissociation is a two step breakup 
process. The projectile in the approaching phase is slowed down by 
the Coulomb field and the valence neutron gets forced oscillation. 
The increasing  long range external force pushes over the projectile 
through the excited states and strongly deforms the orbit of the 
valence neutron.  The projectile survives the collision and surpassing 
the closest approach, released  is trapped into a  highly excited state in 
the continuum. The lifetime of the primary excited state depends on 
the multipolarity of the oscillation. The state of short lifetime
decays prompt in the Coulomb field of the target nucleus, while the long living
state decays delayed, the excited projectile escapes from the Coulomb field
and decays in-flight. The reaction channels are fed  from up downwards 
by the rest of the beam. This is called {\it leakage model of the decay}. 
The intermediate and high energy Coulomb dissociation 
is a sudden process, the beam velocity is larger than the 
velocity of the orbital motion. During the impact the valence neutron stays with 
equal probability in the forward and backward hemisphere of the projectile.
That is the {\it localized valence neutron model}. In the forward hemisphere the
deformation is prolate with dominantly dipole oscillation and single-particle 
excitation, in the backward hemisphere oblate deformation, multipole
oscillation and collective excitation. The dipole excited nucleus decays prompt, 
the multipole delayed. The  secondary decay,  the  final fragments inherit the 
lifetime of the primary decay. So the same fragments can be prompt or delayed 
decay  products, the resonance and the direct breakup processes coexist. 

It must be mentioned in advance that the paper presents the experimental data 
and  in the analysis two simplifications were applied. The events
are taken to Coulomb breakup and the nuclear interaction and interferences
are neglected (those are less than 6\% \cite{Iz2}). The fragment tracks are 
approximated by circles in homogeneous magnetic field. From the observed phenomena 
qualitative statements are taken and an explanation is proposed on semi-classical 
base. These simplifications do not destroy the main message of the experiment, {\it i.e.} 
the inverse population of the reaction channels and that the direct breakup events 
are really  resonance decay events, the projectiles from the low lying  unbound 
resonance state decay prompt in the Coulomb  field of the target nucleus.

\section{Experimental setup}
\label{exp}

The experiment was performed at the Coupled Cyclotron Facility
of the NSCL. The $^8$Li ions were produced by 120~MeV/nucleon  $^{18}$O
beam bombarding  a 2850~mg/cm$^2$ thick  $^9$Be primary target and separated
by the A1900 isotope separator. The $^8$Li beam energy was 40.3 and
69.5~MeV/nucleon (named 40 and 70~MeV/nucleon below) with a dispersion of 
FWHM=1.8~MeV/nucleon. The  reaction target was 56.7~mg/cm$^2$ Pb.   

\begin{figure}[ht] 
\centering
\includegraphics[width=14truecm]{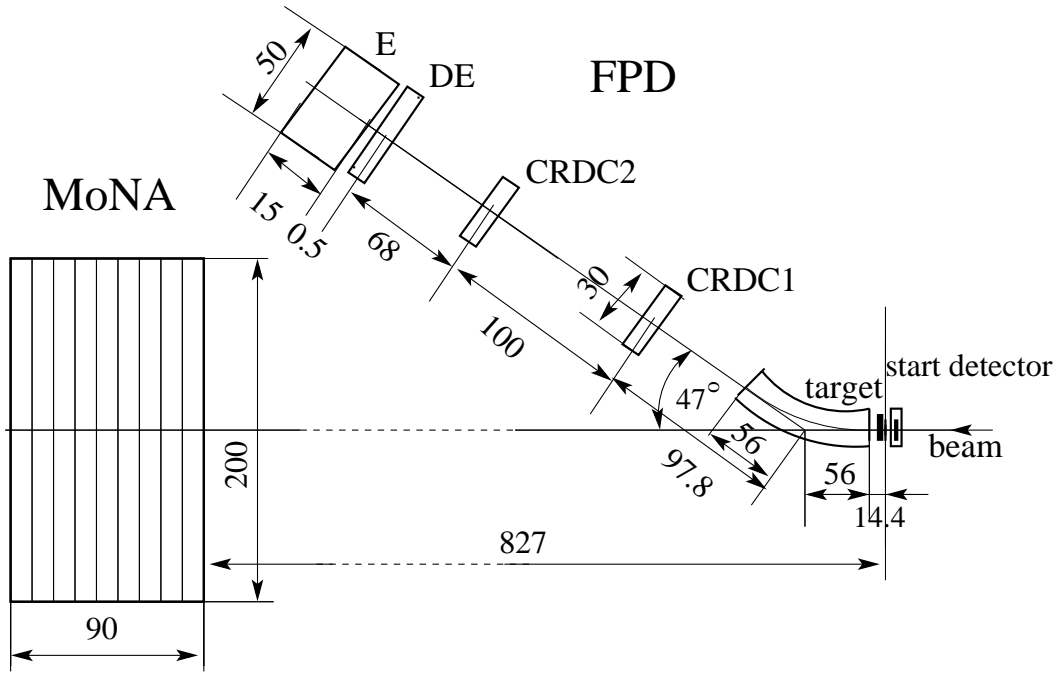}
\caption{MoNA-FPD coincidence spectrometer (distances in cm).}
\label{SP}
\end{figure}

The coincidences of the reaction products were measured by the 
MoNA--FPD (Modular Neutron Array and Focal Plane Detector) 
coincidence spectrometer (fig.~\ref{SP}) \cite{Iz2}. The MoNA neutron detector 
consists of 144 pieces of  $200\times10\times10$~cm$^3$ plastic 
scintillator bars arranged in 9 vertical walls of 16 bars \cite{Ba}. The MoNA
was placed in the beam direction  at zero degree at $D=832$~cm 
from the target. 

The fragments were analyzed by a narrow sweeper magnet \cite{Bi}
followed by the FPD.
The FPD is a telescope of two Cathode Readout  Drift Chamber (CRDC) track 
detectors \cite{Yu}, and a DE, E scintillator pair. The CRDC-s have 
$30\times 30$~cm$^2$ area, 128 horizontal pads. The vertical position
was calculated from the drift time measured on the anode of the CRDC-s. 
The distance of the two CRDC-s was  100~cm at an  angle $47^\circ$ 
from the beam direction in  the laboratory system.
The size of the DE and E plastic scintillators was 
$50\times 50\times 0.5$~ and $50\times 50\times 15$~cm$^3$, respectively.

The magnetic field was set to separate the high velocity fragments
rising from the Coulomb dissociation with $v_F > 0.8~v_{proj}$ for 
fragments A/Z between 2 and 3.

\section{Data processing}
\label{dats}

The parameters used in the analysis  are: the neutron hit position and the time of flight (TOF) 
in the MoNA, and the fragment  hit position and charge  in the CRDC-s, the charge value of the 4 PMT-s of
the  DE and E scintillation detectors, and the fragment TOF in the FPD.  The fragment--neutron 
coincidence events are accepted if the CRDC and scintillator telescopes had valid track, and 
the MoNA bars had valid TOF values at both sides. 

\begin{figure}[ht] 
\centering
\includegraphics[width=14truecm]{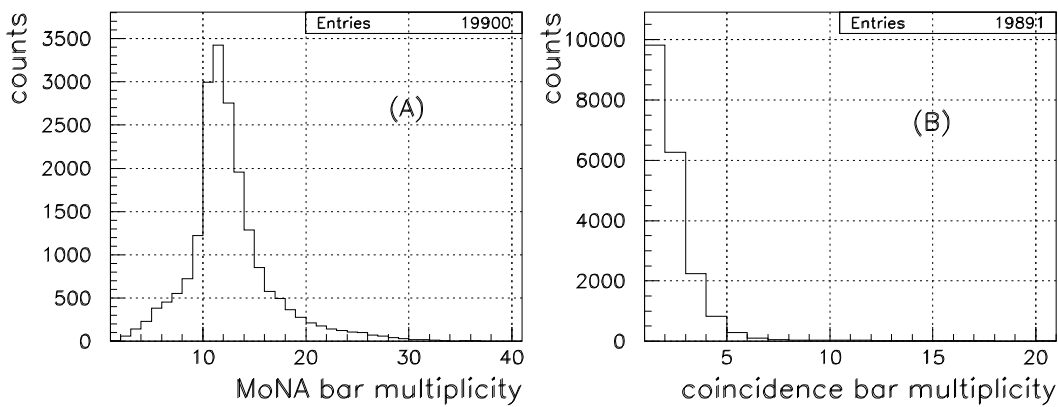}
\caption{MoNA bar multiplicity (a) all bars, (b) valid coincidence bars, at 40~MeV/nucleon.}
\label{MU}
\end{figure}

The FPD and the MoNA detectors started the event taking individually. A common  MoNA  QDC gate 
started the charge integration, while the TDC-s were started individually by the PMT-s and stopped 
by the delayed signal of the start detector. The fragment TOF  was measured in  indirect mode. 
The TDC-s were started by the FPGA clock pulse gated by one of the PMT-s of the DE detector firing 
and stopped by the delayed individual PMT-s of the DE,  E, and the start detector. The events were 
read out and stored if the MoNA--FPD coincidence condition was fulfilled. The coincidence was 
checked by a coincidence gate started by one the DE detector firing. If no coincidence happened, 
a fast clear process (FCL) was started. The MoNA FCL was not working, therefore the coincident 
neutron firing was sitting on several MoNA single data. Because of this error the neutrons of the
good coincidence events have no charge value, the MoNA QDC gate was already closed when the 
coincidence events arrived, and the accidental MoNA TDC-s went to overflow. The average MoNA 
bar multiplicity was 12 (fig.~\ref{MU}(a)), of the selected neutrons 1--3 (b) (see later). 

\begin{figure}[h] 
\centering
\includegraphics[width=7truecm]{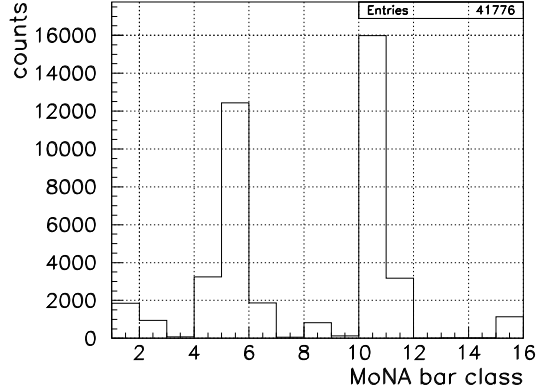}
\caption{MoNA bar class spectrum of the 40 MeV/nucleon coincidence events.}
\label{CL}
\end{figure}

The start scintillator was a 26~mg/cm$^2$ plastic scintillator in front of the target,  
giving  a $\Delta E =2.3$~MeV energy loss. 
The fragments were accepted if both the CRDC and scintillator telescopes have valid values:
\begin{equation}
F=(CRDC1\&CRDC2)\&(DE\&E)  .
\end{equation}
For the true neutron hit selection there was introduced the class parameter of the MoNA hits (fig.~\ref{CL}):
\begin{equation}
CL=1\cdot QL + 2\cdot TL + 4\cdot QR +8\cdot TR.
\end{equation}
Q-s and T-s are the charge and time flags [0,1] of the left and right PMT-s of the MoNA bars.
The valid neutron firings are the neutrons of class $CL=10$ and 15 (TT and QQTT events).
The coincidence events  have 1-3 bars firing (fig.~\ref{MU}(b)). Finally every event
was coincidence event, the proper neutron was selected with high probability only 
it had no valid charge value. The share of the accidental neutron coincidences is 
estimated, it is  a few percent. 

\subsection{Neutron detection}
\label{neu}

The MoNA hit position was determined from the bar sequence 
number --- in the beam and vertical direction --- and horizontally from the time of flight 
difference of the light along the bar: $x=(t_R-t_L)/v_{light}$, where $t$ is the TOF measured
by the PMT-s at the right and left  end of the bar and $v_{light}$ the light velocity in the
scintillator. The $t_n$ TOF of the neutrons is the mean value of the right and left measured 
TOF values: $t_n=(t_R+t_L)/2$.

\begin{figure}[h] 
\centering
\includegraphics[width=7truecm]{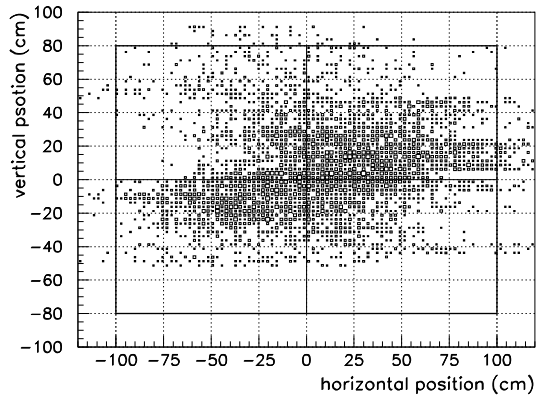}
\caption{MoNA hit map of $^6$Li+n coincidence events at 70~MeV/nucleon.}
\label{xy}
\end{figure}

\begin{figure}[h] 
\centering
\includegraphics[width=14truecm]{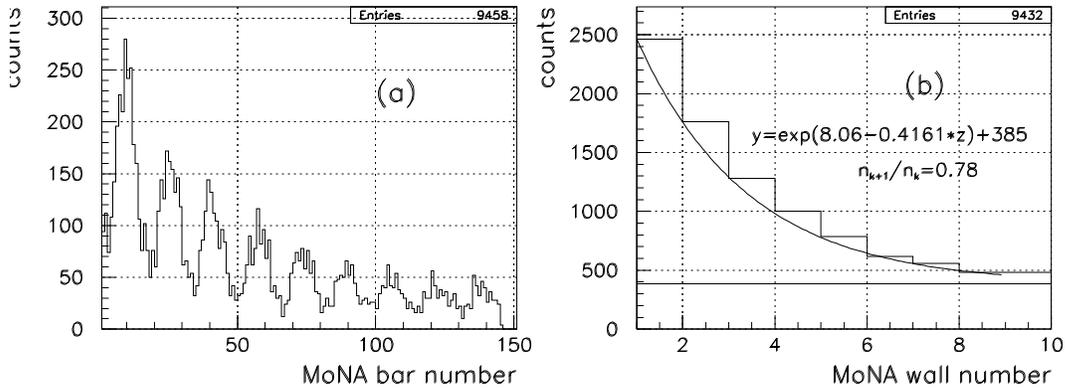}
\caption{(a) MoNA bar frequency of 70~MeV/nucleon coincidence events, (b) attenuation of
the neutrons from the target.}
\label{BA}
\end{figure}

The first MoNA hit of the coincidence events was accepted for the neutron hit. That is the
closest to the target, the direct neutron hit. The MoNA hit map in coincidence with the
$^6$Li fragments is shown in fig.~\ref{xy}. The hits are concentrated around the beam 
direction. The scatter plot has fluctuations. The horizontal structure refers to discrete 
groups of the  $^6$Li+n channel. 
In fig.~\ref{BA}(a) there is shown the spectrum of the
bar frequency of the MoNA. There is a 16-th periodicity with decreasing amplitude. The
particles are flying in the beam direction, they emerge from the target. The horizontal 
distributions are symmetric peaks to the beam direction. 

The particles are neutrons. The area of the peaks decreases exponentially on a constant
(cosmic-rays) background (fig.~\ref{BA}(b)). The detection probability of one bar is 0.6.   

The time calibration of the MoNA TDC channels of the individual PMT-s was performed in two 
steps. The  channels were normalized together by cosmic-rays, and the absolute values 
were calibrated by target $\gamma $-rays. The error of the TOF value was FWHM=0.3~ns
by 75~ns beam TOF (70~MeV/nucleon). The error of the horizontal position calculated 
from the time of flight of the light in the scintillator was $\pm 6$~cm at 827~cm 
target--MoNA distance.  

\subsection{Fragment identification}
\label{Fra}

The published paper \cite{Iz1} focused to the cross section of the 70~MeV/nucleon  $^7$Li+n 
breakup process and did not analyzed the other fragments and the 40~MeV/nucleon data.
The investigation of the reaction mechanisms in the present analysis is based on the whole 
spectrum of the  fragments detected. 

The CRDC parameters are the 
sequence number of the firing pads, the charge measured on the pads and the drift time of the 
anode signal. The horizontal position of the hit was the charge weighted mean of the pads. The 
standard deviation of the horizontal position was 0.4~mm and the angular acceptance
in dispersive $7^{\circ}$. The vertical position was calculated from the  drift time measured 
at the anode. The scintillator telescope measured the $DE$ and $E$ charge produced by 
the fragments and the TOF values. 

\begin{figure}[h] 
\centering\includegraphics[width=6truecm]{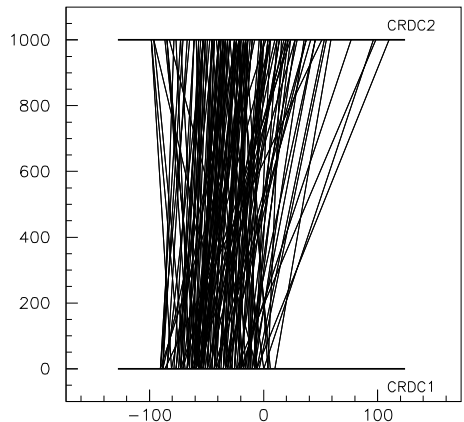}
\caption{Asymptotic tracks in the CRDC telescope of $^6$Li fragments at 40~MeV/nucleon.} 
\label{trk}
\end{figure}

The fragment identification is performed in two steps. First the fragments are classified into
groups and then the Z and A are assigned to the groups based on the energy and the
correlation of the fragment and neutron velocities. The grouping of the fragments was performed
only by the direct measured parameters: CRDC hits, scintillator DE, E charges and the DE TDC
(TOF) values. The asymptotic tracks are shown in fig.~\ref{trk} of the $^6$Li  fragments --- 
between the CRDC-s ---  at 70~MeV/nucleon. Two kinds of particle tracks can be observed. There
are parallel, close to   perpendicular and crossing  tracks: the resonance and direct breakup events.

\begin{figure}[ht]  
\centering
\includegraphics[width=7truecm]{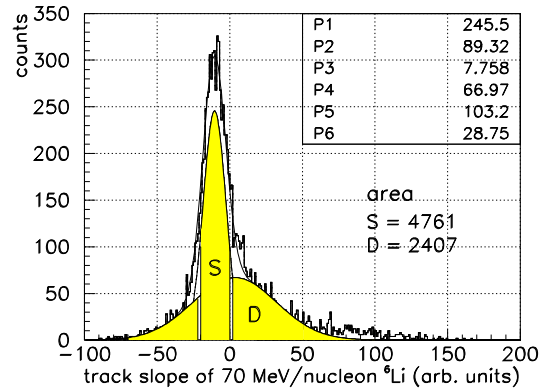}
\caption{Slope of the asymptotic tracks of $^6$Li fragments at 70~MeV/nucleon energy fitted by two Gaussian functions.} 
\label{SL}
\end{figure}

From the CRDC telescope coordinates one can calculate the slope of the asymptotic tracks in 
the CRDC-s frame as $SL=(w2-w1)/d$, where $w$-s are the hit coordinates in mm 
($w=\pm 150$~mm) and $d=1000$~mm the distance of the CRDC-s (fig.~\ref{SL}). 
The error of the track slope is 0.1\%. In the plot of the $^6$Li track
 slope at 70~MeV/nucleon, there is a narrow peak of the resonance decay events  
and a broad one of the direct breakup events. The spectrum is fitted by two Gaussian 
functions (S) and (D) (the parameters are: amplitude, position, and standard deviation 
of the peaks). The ratio of the two components is 50--50\%.

The path of flight of the fragments in the magnet was approximated  fitting a circular 
orbit across the target point and  the outgoing track in a constant magnetic field. 
Therefore the fragment path is a rough approximation, but at small scale it proved to 
be admissible (the error of the path of flight $\Delta L/L<1$\%). 
This simplified analysis is satisfactory for relative and qualitative 
statements. The reaction channel can be identified by the fragment data, while the 
neutron data can give information about the decay  process.

\begin{figure}[h] 
\centering
\includegraphics[width=14truecm]{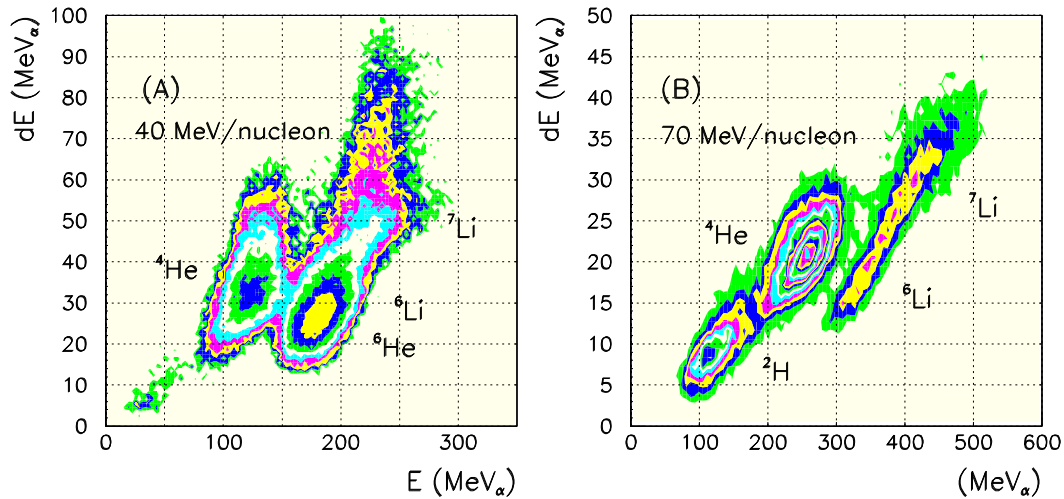}
\caption{Fragment $dE~vs~E$ contour plots (A) at 40 and (B) at 70~MeV/nucleon.
 The axises are calibrated to the $^4$He.}
\label{EdE}
\end{figure}

The fragment energy spectra of the scintillator telescope are shown in the 
contour plots $dE~vs~E$ in fig.~\ref{EdE} at 40 and 70 MeV/nucleon energies. 
The charge values of the DE and E scintillators are calibrated to the $^4$He group 
calculated with the program  DONNA. For the other ions the $dE_\alpha $ 
and $E_\alpha $ values are approximations neglecting the difference of the light output
for different ions and the non linearity of the photomultipliers. Several isotope 
groups can be observed at both energies, but the plots are not appropriate for 
fragment separation.


\begin{figure}[h] 
\centering
\includegraphics[width=14truecm]{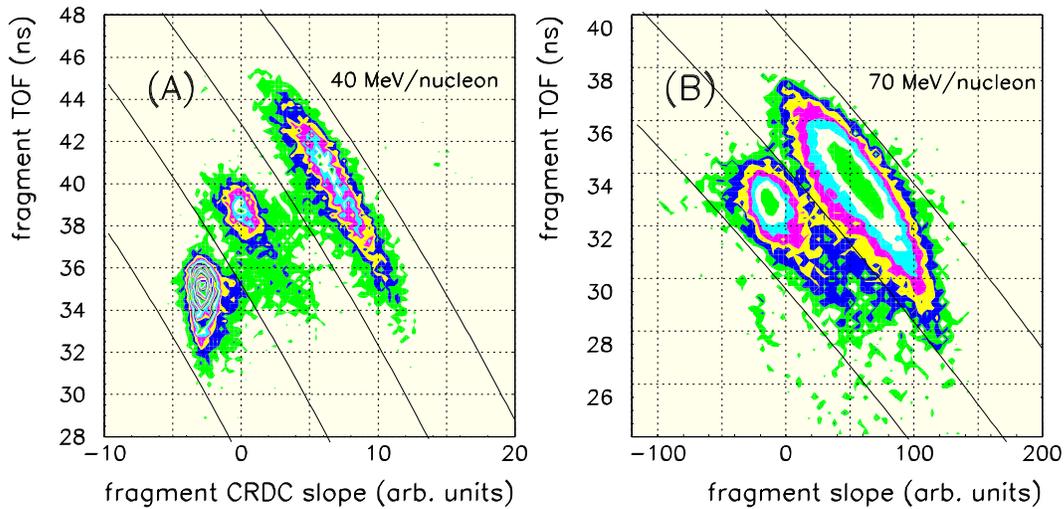}
\caption{Fragment TOF~{\it vs}~slope of the asymptotic tracks (A) at 40 and (B) at
70~MeV/nucleon. The lines are examples of iso-separation lines.}
\label{SLtof}
\end{figure}

The kinematic behavior of the fragments can be studied in the velocity map of 
the fragment  TOF {\it vs} PoF fragment path of flight. As the fragment path is 
a rough approximation, instead of the velocity the  TOF {\it vs} SL contour plot
is used for classification of the events  (fig.~\ref{SLtof}), which are direct 
measured parameters.


At 40~MeV/nucleon (A) three and at 70~MeV/nucleon (B) two 
groups are well separated. For the separation the groups one can construct 
concentric circles which fit to the curvature of the groups. The events can be  
characterized by the radius of a circle crossing the point of the event and 
concentric with the cutting ones. The separation radius parameter $R_s$ is an 
arbitrary pseudo parameter: the distance of the point from  circle crossing 
the origin of the plot.
\begin{equation}
 R_s=\sqrt{(x-x_0)^2+(y-y_0)^2}-\sqrt{x_0^2+y_0^2},
\end{equation}
\noindent
where $x$ and $y$ are the $SL$ slope and $TOF$ coordinates and $x_0$ and $y_0$ are the 
coordinates of the center of the concentric circles.  
The separation radius is ordered to the event. It is  an
empirical parameter for separation the fragments one by one
according to their kinematic parameters: momentum, scattering angle, and charge.
The groups are well separated. They can be cut with a definite 
efficiency. Some iso-separation lines are  plotted to show the quality of 
the separation. 


\begin{figure}[h] 
\centering
\includegraphics[width=14truecm]{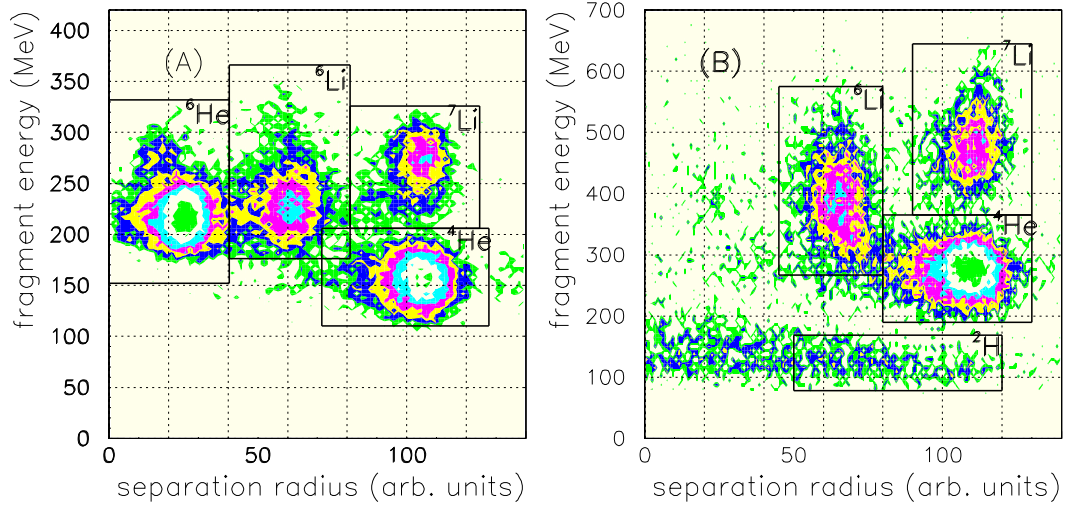}
\caption{Fragment energy {\it vs} separation radius at (A) 40 and (B)  
70~MeV/nucleon. The boxes are the separating intervals.}
\label{SReE}
\end{figure}

\begin{figure*}[th]  
\centering
\includegraphics[width=15truecm]{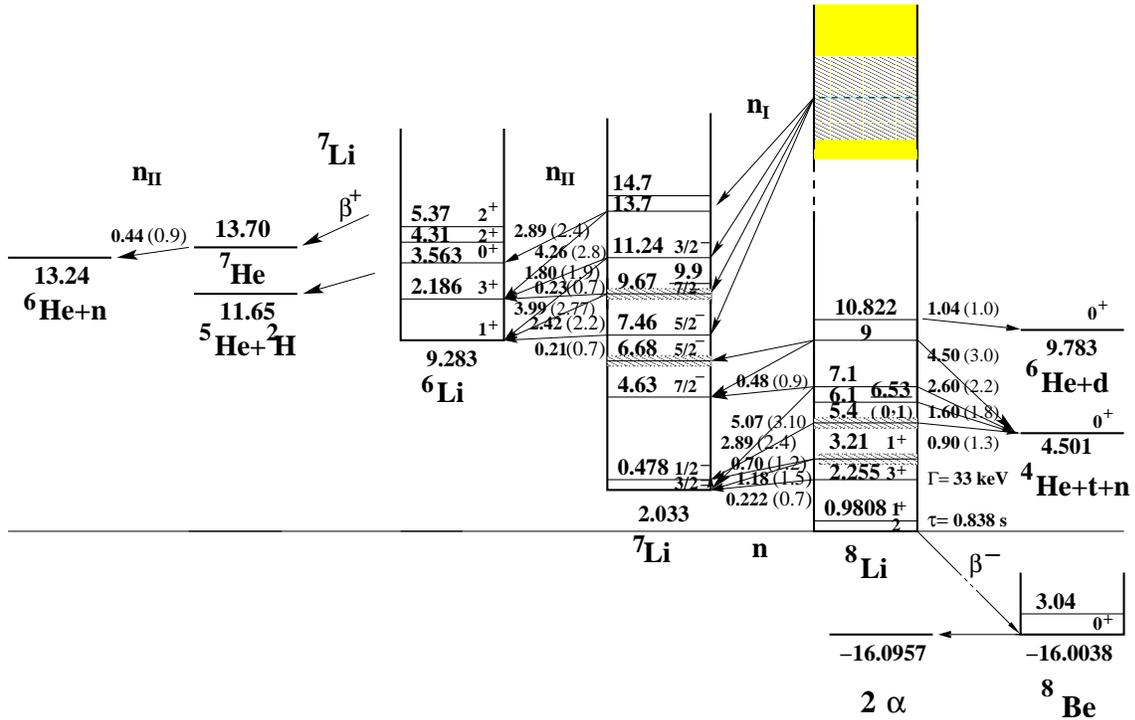}
\caption{$^8$Li decay scheme \cite{AS1}. The neutron decay energy and velocity 
(in brackets) are in MeV and cm/ns.}
\label{Li8}
\end{figure*}

The sum energy of the fragment $E=E_\alpha +dE_\alpha $  is an appropriate  
parameter for the fragment identification.  The $E~ vs ~R_s$  plots give 
consummate  separation of the fragments (fig.~\ref{SReE}). 
The separation of the groups at higher energy is better.
In the plots four groups can be separated  at 40 and 70~MeV/nucleon. The boxes  
are the limits of the interval tests.  There are a triple group and a fourth one. 
The fourth groups are  different  corresponding to the two magnetic field settings.  
The mass number can be read out from the  plots, from the energy of the ions. 
The groups are  at $A=4,6,7$ times 40, and $A=2,4,6,7$ times 70~MeV energy.
Supposing that the events are neutron--fragment coincidences, a charge can be 
assigned to the groups. The identification is verified by the velocity 
correlations  of the fragment and the neutron (see Sect. 7.3). 
At 40~MeV/nucleon $^7$Li, $^6$Li, $^6$He,  $^4$He fragments, 
at 70~MeV/nucleon $^7$Li, $^6$Li, $^4$He, and $^2$H  can be identified. The 
suggested decay scheme  of the $^8$Li is shown in fig.~\ref{Li8} \cite{AS1,AS2}.  
The possible observed transitions are signed with the energy and velocity values. 
The latter are in brackets in cm/ns unit.

This fragment identification method can be useful during the data taking, on-line, 
for it does not need special calibration and time consuming calculations.


\section{Inverse population ratio} 
\label{inv}

The threshold of the reaction channels is rather high. 
The $^4$He+t+n channel opens above 4.5, the $^6$Li channel above 9.28, 
the $^2$H channel above 11.65, the $^6$He+n channel 13.7~MeV.  
The events of the $^6$He+d channel are accidental neutron coincidence events.  
The $^4$He+t+n is a three particle final state, the spectra are broader. 
The $^6$Li+2n, the $^5$He+$^2$H+n are two-step,  and the $^6$He$+\beta +$2n 
reaction is three-step process. 
The relative yields of the fragments are listed in table \ref{tab1}. 
read from the plots fig.~\ref{SReE}. The share of the  $^7$Li+n  channel is 
14 and 16\% at 40 and 70~MeV/nucleon. The errors are the statistical errors.

 At 40~MeV/nucleon  the dominating reaction channel is the 
$^6$He with 36\%, next is the $^4$He 27\%, and  the $^6$Li+2n
channel with 23\%. It has to be noticed that the $^6$He+d channel is
overrepresented because of the accidental single-MoNA--fragment
coincidences. The share of the $^6$Li  and the $^2$H is reduced 
because of the  two-step decay, and the small geometric 
efficiency of the MoNA for the high energy neutrons (about 20\%). 
The real order of the branching is probably  $^6$He, $^2$H, $^6$Li, $^4$He, 
and $^7$Li. From the intensity ratios at both energies the conclusion 
is plausible that the $^8$Li is excited high into the giant resonance 
region and decaying feeds the different reaction channels. The 
larger the energy of the secondary excited state, the larger the decay 
probability through that channel. That is called {\it inverse population 
ratio}. The reaction channels are fed from up to downwards by the rest 
of the beam particles.

\begin{table}[!h]

\caption{Relative yield of fragments.}
\label{tab1}
\vskip 1pc
\centering
\begin{tabular}{@{}cccccccc}
\hline\noalign{\smallskip}
beam energy&fragment&threshold&counts&partition\\
 MeV/nucleon& &MeV & & \%\\
\noalign{\smallskip}\hline \noalign{\smallskip}
40 &  $^6$Li& 9.28&  9105&  23$\pm 0.24$  &\\
40 &  $^7$Li& 2.03&  5537&  14 $\pm 0.19$&\\
40 &  $^4$He& 4.50& 11000&  27 $\pm 0.25$&\\
40 &  $^6$He+d& 9.78&  13604&  34 $\pm 0.29$&\\
40 &  $^6$He+n& 13.7&  950&  2 $\pm 0.06$&\\
\hline\noalign{\smallskip}
70 &  $^6$Li& 9.28&  9458&  24$\pm 0.25$ &\\
70 &  $^7$Li& 2.03&  6896&  16$\pm 0.19$ &\\
70 &  $^4$He& 4.50& 21114&  52 $\pm 0.36$&\\
70 &  $^2$H& 11.65&  3076&   8 $\pm 0.14$&\\
\noalign{\smallskip}\hline 
\end{tabular}

\end{table}

\section{Reaction mechanisms}
\label{mech}

For studying the reaction mechanism of the Coulomb dissociation processes
the best example is the $^6$Li+2n channel because of the best statistics.
The reaction is a sequential two-step decay: $^8$Li$^*\rightarrow^7$Li$^*$+n$_I$
and $^7$Li$^*\rightarrow^6$Li+n$_{II}$. The neutrons n$_{II}$ in the 2$^{nd}$ 
step of the dissociation are low energy ones.  

Figure~\ref{SL} shows the slope spectrum of the asymptotic tracks of the  $^6$Li in 
the CRDC  frame, at 70~MeV/nucleon. It is fitted by two Gaussian functions. 
The narrow  peak is the in-flight decaying  long-lived resonance decay (group S) 
and the tail the direct breakup process (group D). The  two reaction 
mechanisms can be separated  confidently. 


\begin{figure*}[t]  
\centering
\includegraphics[width=15truecm]{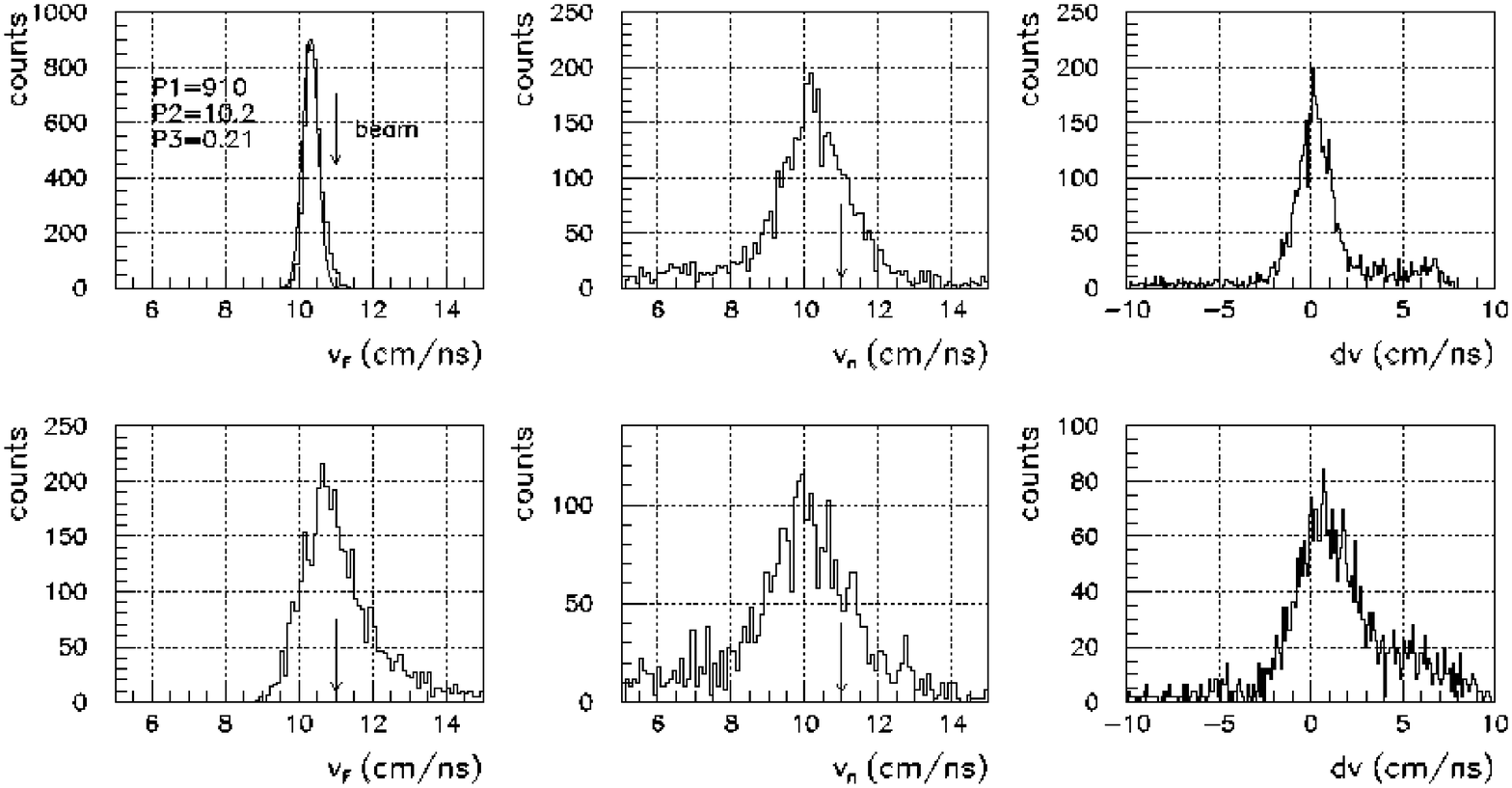}
\caption{Fragment and neutron velocities of $^6$Li+n$_{II}$ coincidences of 
70 MeV/nucleon  $^6$Li+2n reaction channel. Resonance (upper) and direct breakup 
events (lower row). The left column is the fragment, middle the neutron velocities and the 
right one $dv=v_F-v_n$.  The fit is a Gaussian distribution.}
\label{vrd}
\end{figure*}

To verify the two reaction mechanisms  in fig.~\ref{vrd} there are plotted  
the $v_F$ fragment and $v_n$ neutron velocities, and the $dv=v_F-v_n$ 
velocity differences in cm/ns of the 70~MeV/nucleon 
$^7$Li$^*\rightarrow^6$Li+n$_{II}$  reaction. 
The upper row shows the resonance (group S)  and  the 
lower row the direct breakup events (group D) gated by the track slope 
parameter (fig.~\ref{SL}). The beam velocity  is 11.0 cm/ns.  
Left column is the fragment, the middle  the neutron
velocities and the right one the $dv$ velocity difference. 

The fragment velocity peak of the escaped and in-flight decayed resonance 
group is narrow. The contribution of the D component in the resonance 
section of the velocity spectra is less than 5\%. The standard 
deviation is $\sigma =0.21$~cm/ns, which corresponds to 23~keV.
The velocity shift is $v_s=-0.8$~cm/ns.
The velocity difference $dv$ is a narrow peak around zero. 
The neutron velocity has a peak with sequential decay  
structure, symmetric to the fragment peak. It will be discussed in Sec.~6.1. 
The velocity spectra of the  direct breakup  events are broader  
because the position of the decay in the Coulomb field is not definite. 
The post-breakup acceleration of the fragments produces a large 
velocity tail above the beam velocity. The neutron velocity spectrum has
similar structure as the resonance group but shifted down (about 0.1~cm/ns)
and broader. This supports the  conjecture that the direct breakup is 
also a resonance-like, but prompt sequential decay process of a source 
braked down inside the Coulomb field.  

Similar plots can be produced  from the other reaction channels at  
40 and 70 MeV/nucleon.

\section{Delayed, resonance decay}
\label{res}

The neutron velocity spectrum of the resonance decay events (group S)  is a sequential 
neutron decay spectrum. The fully re-accelerated excited projectile decays 
in-flight emitting  a discrete energy neutron (fig.~\ref{vrd}).
The reaction products are kinematic focused. The velocity spectrum of  discrete 
resonance decay particles emitted into $4\pi $ solid angle  is a trapezoid function. 
If the transverse momentum is larger than a certain value the particles will miss 
the detectors, fall  out of their solid angle. The middle of the trapezoid spectrum
will be cut out and the spectrum consists from  a symmetric  peak pair of the forward 
and backward emitted neutrons. The particles are detected in a cone of $\Phi _0$ 
limiting angle. The limiting angle for the decay momentum $q$ in the laboratory 
frame is: 
\begin{equation}
\phi _0 =\arcsin(\frac{X}{\sqrt{D^2+X^2}} \cdot \frac{p}{q})\pm \arctan(\frac{X}{D}) ,
\end{equation}
where $D$ is the target--detector distance, $X$  is the half of the detector size, $p$ 
is the momentum of the beam velocity particle in the moment of the decay, taking into 
account the deceleration and the excitation velocity  loss. The cutting results in a gap 
at the middle of the spectrum. However, the velocity and angular dispersion of 
the beam expands the velocity region, that will be cut but at the same time it
will fill the gap and can form a pileup peak in the middle overlapping region.
The limiting velocity according to equ.~(4) is $v_c=0.1193 v_0$~cm/ns.

\subsection{$^6$Li+2n channel}
\label{li61}

The velocity spectrum of the $^6$Li  group S  has a resonance character (fig.~\ref{vrd}). 
The narrow peak and the correlation of the $^6$Li--neutron velocities
(Sect. 7.3.1) verify that the fragments are from the decay of a  resonance state  
$^7$Li$^*\rightarrow ^6$Li+n$_{II}$.
The threshold energy of the reaction channel is 9.28~MeV. One has to suppose that 
the $^8$Li$^*$  fell into a long living resonance state  high in the continuum, 
re-accelerated escaped from the Coulomb field of the target nucleus and  emitting a 
neutron fell into a highly excited resonance state  of the $^7$Li$^*$ 
(above 7.46~MeV), which decayed in-flight into $^6$Li+n$_{II}$. 
The decay  energy  can be  estimated from the $\Delta E_n$ neutron energy increment 
corrected for the recoil on an $A_F$ mass fragment $E_D=\Delta E_n (1+1/A_F)$. 

\begin{figure}[h]
\centering
\includegraphics[width=7truecm]{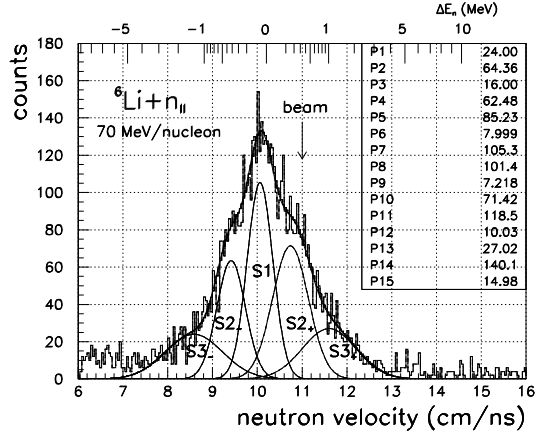}
\caption{Neutron velocity  spectrum of  $^6$Li+n$_{II}$ resonance decay events at 70~MeV/nucleon cut by track slope gate S. The S1--S3 are Gaussian function fits. The fits are in 0.04~cm/ns units.}
\label{vn76}
\end{figure}

The neutron velocity spectrum of the 70~MeV/nucleon $^6$Li+n$_{II}$ resonance events 
is shown in fig.~\ref{vn76}. It consists of a single (S1) and two  double peaks 
(S2 and S3) symmetric to the single one. The peaks are fitted by Gaussian distributions.
The symmetry axis is shifted down by $v_s=-0.8$~cm/ns related to 
the beam velocity. The resonance decay particles have a definite velocity deficit. 


The distances of the  S2 and S3 peaks from the symmetry axis are 
$\Delta v_n=(v_+-v_-)/2=\pm 0.61$ and 1.5~cm/ns. These velocities 
are compatible with the decay scheme of the $^7$Li \cite{AS1} (fig.\ref{Li8}). 
The decay velocity of the totally separated peaks (S3) are underestimated. 
The spectrum of the group is cut, and the decay velocity is the extreme forward--backward 
velocity value, however, the fit  gives the center of the  sections. 
The standard deviation of the peak has to be added. With $\sigma =0.62$~cm/ns 
the neutron velocity shift is $\Delta v_{nS3}=\pm 2.12$~cm/ns. The S2 groups 
are in the cone of the solid angle of the MoNA, but the beam dispersion 
expands the dynamics of the neutrons and the wings of the distributions
are cut outside and give a pileup peak in the middle. The central peak S1 with 
$v_D=0\pm 0.35$~cm/ns  has no equivalent transition.  It must be  the 
pileup peak of the  S2 transition. The recoil corrected decay 
velocities are $v_D=(1+1/A_F)\cdot \Delta v_n=0.71$ and 2.47~cm/ns. The 9.67~MeV state 
of the $^7$Li decays to the first excited and the ground state of the $^6$Li with 
$v_D=0.7$ and 2.2~cm/ns neutron emission. 

\begin{figure}[ht]
\centering
\includegraphics[width=14truecm]{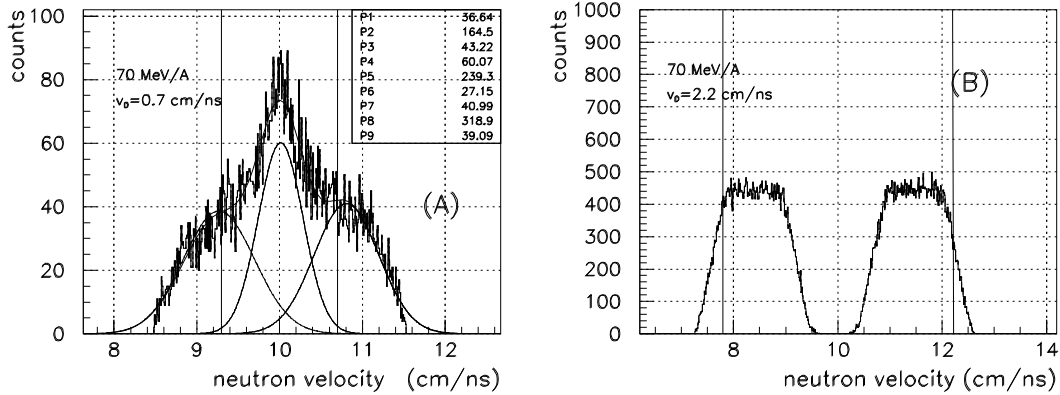}
\caption{Monte Carlo simulation of the neutron velocity spectra of the 0.21~MeV ($v_D=0.7$cm/ns) (A) and the 2.42~MeV  ($v_D=2.2$~cm/ns) transitions (B) at 70~MeV/nucleon. The vertical lines show the $v_D$  decay velocity.} 
\label{MC1}
\end{figure}

With Monte Carlo simulation the 30\% overshoot can be produced by 20\% overlap 
of the  forward--backward sections. The plots in fig.~\ref{MC1}(A) and (B) simulate 
the neutron velocity spectra measured in coincidence with the $^6$Li fragments at 
70~MeV/nucleon energy. The 0.21~MeV transition is simulated with $v_D=0.7$~cm/ns 
decay velocity. The limiting momentum $q$ is 47~MeV/c ($E_D=1.2$~MeV) at 70~MeV/nucleon.
Therefore the  decay velocity is inside the acceptance of the MoNA and FPD, 
however, the effective velocity range is extended  by the  velocity and the angular 
dispersion of the beam.  The parameters of the simulation  
are the velocities corresponding to the dispersion of the  beam energy ($1.8$~MeV/nucleon),
the angular dispersion (0.06~rad),  the velocity resolution (0.04~cm/ns)
and the limiting momentum (eq. 4). The limiting neutron velocity is fitted to 
reproduce the spectra at both energies: $v_c=0.76$  and 0.81~cm/ns at 40 and 
70~MeV/nucleon. The neutrons of the higher decay energy events give separated peaks. 
Figure \ref{MC1}(B) shows the simulation of the 2.42~MeV transition. Please notice 
that the peaks are not Gaussian functions and the decay velocity is at the edge 
of the peaks. 


The other fragments have also resonance decay events. Those can be selected by the 
track slope parameter, or the fragment--neutron velocity difference, however, the
separation of the two mechanisms is less clear. 

\section{Prompt, direct breakup}
\label{prmt}

In the direct breakup process the dissociation happens in the 
Coulomb field of the target nucleus. The (A,Z) projectile is braked down 
in the strong repulsing field, emits a particle and the residual
(A-a,Z-z) fragment will be accelerated. The energy stored in the 
Coulomb potential accelerates the fragment. The specific energy loss and 
gain of a charged particle is proportional to the path taken with 
different  Z/A.  The relative specific energy increment  of the 
fragment  in the Coulomb field is $\varepsilon =(Z_F/A_F-Z/A)/(Z/A)$.
The relative velocity increment of the fragment, the  post-breakup 
acceleration, is the square root of the specific energy excess 
$\Delta v_F/\Delta v_n=-\sqrt{\varepsilon}$. Here it is assumed 
that the velocity decrease of the neutron is equal with the velocity 
loss of the source, {\it i.e.} of the  projectile in the moment of 
the decay. The  post-breakup acceleration is suitable 
for the fragment identification of a fragment group with known mass number.  

\subsection{Neutron velocity of prompt breakup events}
\label{n-prmt}

\begin{figure}[h] 
\centering
\includegraphics[width=14truecm]{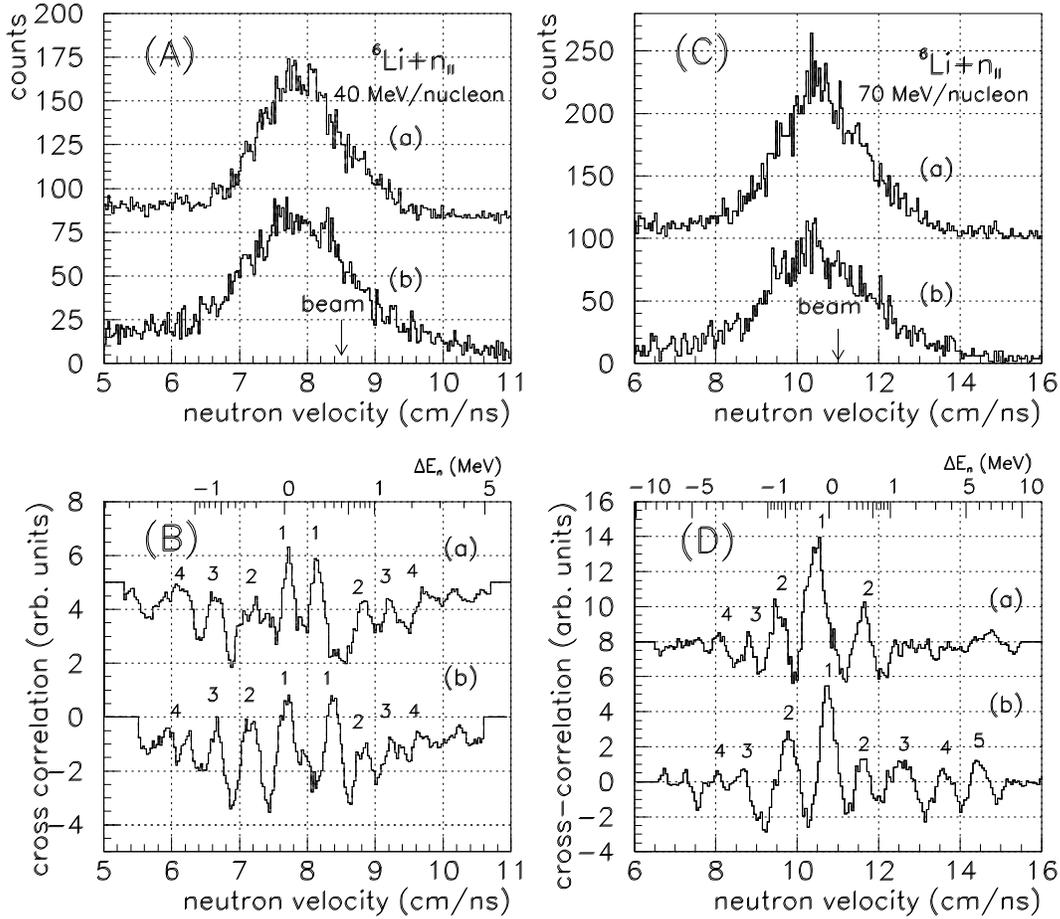}
\caption{Neutron velocity spectra of the $^6$Li+n$_{II}$ channel  (A) at 40 and (C) 70~MeV/nucleon energies. (B) and (D) are their  cross-correlation functions by a Gaussian searching function with $\sigma =0.15$ and 0.25~cm/ns standard deviation, respectively. (a) is the resonance and (b)  the direct breakup components.}
\label{cvn}
\end{figure}

The structure of the neutron velocity spectra 
can be studied by their cross-correlation function with a searching form
function  \cite{Co}.  In fig.~\ref{cvn} there are shown the neutron velocity 
spectra and their cross-correlation with a Gaussian searching function
of the  $^6$Li+n resonance and prompt decay events at 40 and 70~MeV/nucleon energy.
The cross-correlation code calculates the difference between the channel 
content and the mean value in a sliding surrounding window weighted by the form function 
of the searched peaks.  The statistical error, the square root of the mean is added 
to the mean value. (The cross-correlation method is used in the automatic 
spectrum analysis codes.) The standard deviation  of the  form function was $\sigma =0.15$ 
and 0.25~cm/ns at 40 (A) and 70~MeV/nucleon (B), and the searching window was $\pm 2\sigma $ wide.  
The cross-correlation function is:
\begin{eqnarray}
C(i)=\sum_{j=-k}^k G(j)\cdot\left(F(i+j) -(\overline{F(i)}+\sqrt{\overline{F(i)}}~)\right)\\  j=-k,.,0,.,k\nonumber
\end{eqnarray}
\begin{eqnarray}
G(j)=\frac{1}{\displaystyle{\sqrt{2\pi }}{\displaystyle\sigma }} \exp{(-\frac{j^2}{2\sigma ^2})}\\ \overline{F(i)}=\frac{1}{2k+1}\sum_{j=-k}^k F(i+j) ,
\end{eqnarray}
where $C$ is the cross-correlation function of the $F$ velocity spectrum with a normalized
Gaussian searching function $G$. 

The peak structure of the cross-correlation functions verifies that the neutrons 
are emitted from excited states with discrete excitation energy.  
The cross-correlation functions of the delayed  resonance (a) and the prompt, direct 
decay spectra  (b) have identical structure at both energies, although they originated 
from different decay processes. This refers to common origin.
The peaks in the cross-correlation plots are labeled with numbers. The spectrum of 
the 70~MeV/nucleon events (D) has a strong central peak (No.~1.) and
some symmetric satellite peak  pairs. At 40~MeV/nucleon (B)  the central peak is split 
because of the less kinetic focusing (cut forwards/backwards).  
The beam velocities are 8.55 and  10.99~cm/ns. The symmetry axises,
(peaks No.~1), are shifted down by $v_s=-0.8$~cm/ns relative to the beam velocity. 
At 40~MeV/nucleon the  resonance and direct breakup spectra are rather similar because 
of the separation of the two decay  mechanisms is not sharp.   

The velocity deficit corresponds to the kinetic energy spent to the  excitation of a primary 
high energy state $^8$Li$^*$ in the continuum $v_s=v_x$. At 70~MeV/nucleon  the impact parameter 
was $b=20$~fm \cite{Iz2}. Neglecting the perpendicular displacement of the track the Coulomb potential 
of a Li ion at 20~fm from the Pb target nucleus is 18~MeV. The $^8$Li is a dripline  nucleus, 
the $^7$Li core can be taken as structure less particle which binds the valence neutron weakly. 
The $v_s=-0.8$~cm/ns velocity shift  of the escaped projectile, the resonance decay, 
corresponds to 24.9~MeV/c momentum transfer to the continuum particles, in agreement 
with the plot  shown in ref.~\cite{BB}. In case of single particle excitation the
deformation and the excitation energy is concentrated to the valence neutron which takes
199~MeV/c momentum  or 21~MeV excitation energy.

The down shifted symmetry axes is the velocity of the fully re-accelerated, binding energy
corrected projectile $v_0=v_{beam}-v_x$. It was found, that  $v_0$ is common for the $^6$Li, 
$^7$Li, and $^4$He reaction channels $v_0=7.8$ and 10.2~cm/ns at 
40 and 70~MeV/nucleon, respectively.   The peaks can be co-ordinated 
to each other in the prompt and resonance  cross-correlation spectra.  The  position of 
the peaks in the cross-correlation spectrum  depends on the environment. The large peak 
shifts the side ones. The  error can be in the order of $\pm 0.2$~cm/ns.  The measured, 
recoil charged decay velocity ($v_D'=A/(A+1)\cdot v_D$) can be read out from 
fig.~\ref{cvn}.  The average of the peak pairs:  $v_D'=(v_+-v_-)/2$, 
where $v_\pm$ are the position of the forward, backward peaks.  At 70~MeV/nucleon  
the peak No.~1 is a single peak in the center.  The recoil corrected  decay velocities of 
the peaks are: $v_{D2}=0.8$, $v_{D3}=1.8$, and  $v_{D4}=2.6$~cm/ns. These peaks  can be 
identified with the $^7$Li$^* \rightarrow^6$Li+n decay. The transitions from the three 
levels above the threshold energy of the $^7$Li 7.46, 9.67, and 11.24~MeV  to the ground 
and two excited states of the $^6$Li correspond  to the measured peaks with 0.7, 1.9, 2.2, or 
2.4~cm/ns decay velocities. The peak No.~1.  has no corresponding transition. 
It must be the  pileup peak of the  peaks No.~2 at $v_0$, because of the  dispersion 
of the beam  velocity.  At 40~MeV/nucleon  the peak No.~3.
(1.4~cm/ns) cannot be identified in the decay scheme, but the $v_{D2}=0.8$, 
$v_{D4}=1.8 \pm 0.2$~cm/ns transitions can be verified. 

\begin{figure}[!ht] 
\centering
\includegraphics[width=14truecm]{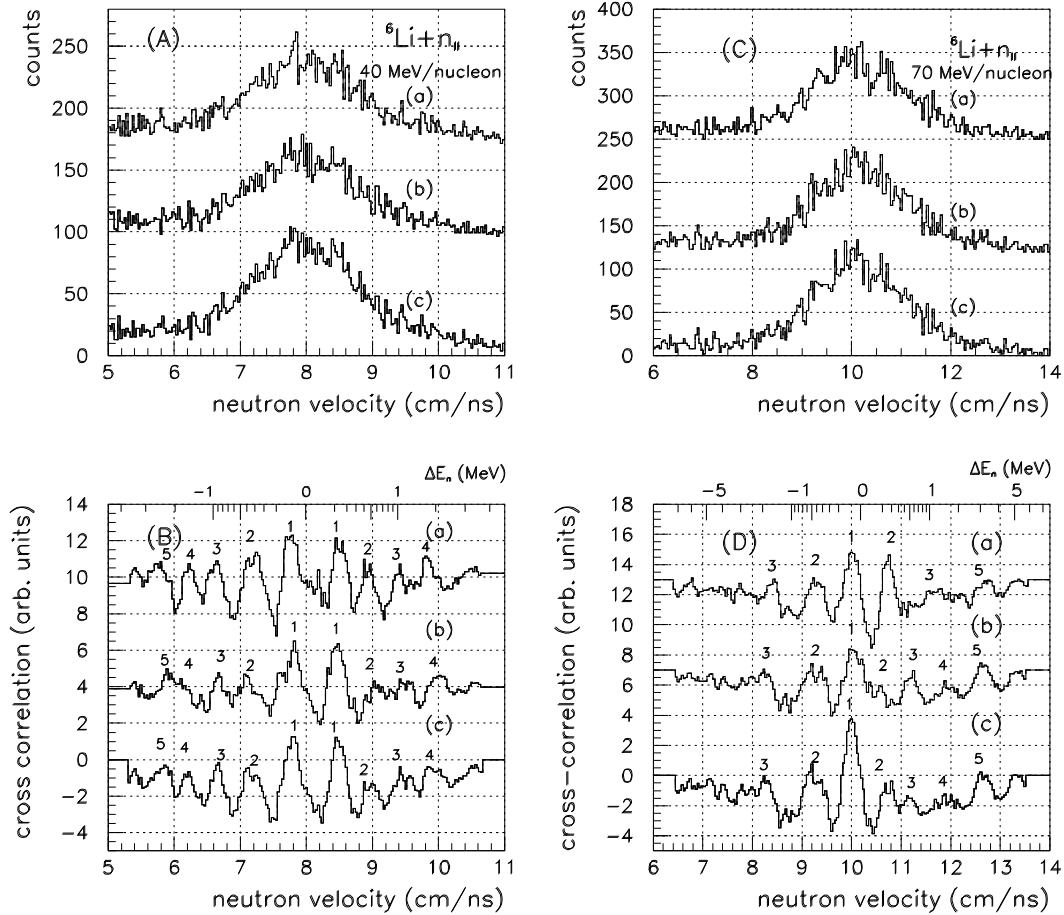}

\caption{Neutron velocity spectra of the $^6$Li direct breakup events at 40 (A) and 70~MeV/nucleon  energy (C). (B) and (D) are their  cross-correlation functions by a Gaussian searching function with $\sigma =0.15$ and 0.25~cm/ns standard deviation, respectively. The curves (a), (b), and (c) are the two halves of the events, and the full statistics.}
\label{cvnB}
\end{figure}

Regarding the simplified analysis and the poor statistics the question occurs 
whether are the peaks  true or accidental ones? 
In order to verify that the structure of the neutron velocity spectrum is true, it corresponds 
to the  decay, to the Coulomb breakup process, the data of the direct breakup events are 
cut into two halves and their cross-correlation functions are compared in fig.~\ref{cvnB} at both 
energies. Most of the peaks can be identified in both halves and at both energies. 
The comparison of the two halves of the events verifies that the peaks are real peaks. 

\begin{figure}[ht] 
\centering
\includegraphics[width=14truecm]{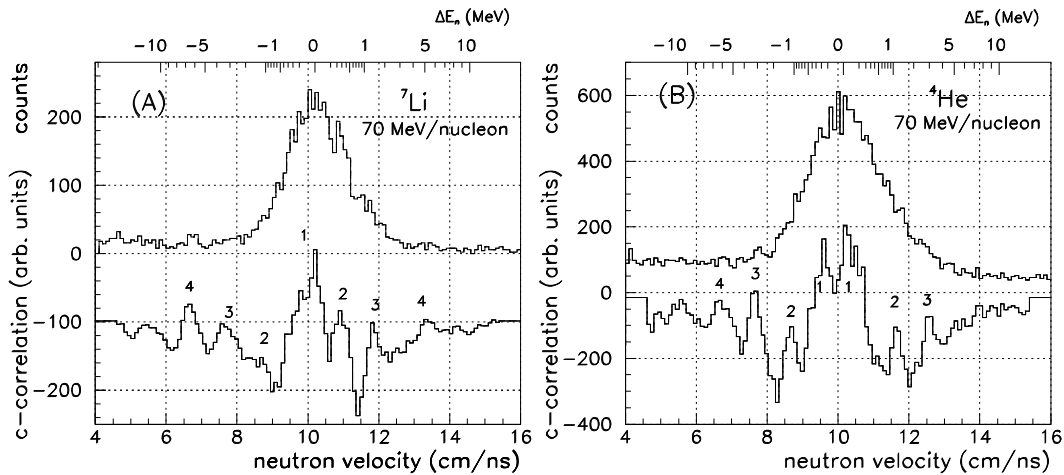}
\caption{Neutron velocity spectra and the cross-correlation function of 70~MeV/nucleon $^7$Li+n and $^4$He+t+n reaction channels.} 
\label{cvLiHe}
\end{figure}

The cross-correlation functions of the  neutron velocity spectra in the $^7$Li and $^4$He 
channels are shown in  fig.~\ref{cvLiHe}. The velocity shift is  $v_s=-0.8$~cm/ns.
The peaks No.~1. can be pileup peaks at  0  and
0.35~cm/ns, but the peaks in the $^7$Li channel at 1.10, 2.05, and 3.25~cm/ns can correspond 
to the 1.2--1.5, 2.4, and 3.1~cm/ns decay velocities.  Similarly the
1.45, 2.40, and 3.30~cm/ns peaks of the $^4$He channel can be identified  as the 
transitions with 1.3, 2.2, and 3.0~cm/ns decay velocities. 

Summarizing please notice that the cross-correlation analysis gives valuable information about 
the behavior of a statistical mass, about its fluctuation. It was found that the structure  
of the velocity spectra of the neutrons from the prompt and delayed decay are identical, 
we can tell that in both mechanisms there are events from the same  transitions.
The two decay modes, the resonance and direct breakup  must differ  in  the lifetime
of the primary excited states. The  delayed, resonance  decay neutrons are low energy 
n$_{II}$ neutrons but the prompt decay events have high decay energy primary neutrons 
too (fig.~\ref{cvn}). In  40~MeV/nucleon the separation of the two mechanisms is worse
than at 70~MeV/nucleon. The separated resonance decay events 
contain 34\% prompt decayed events, therefore the resonance and direct breakup 
spectra  are rather similar. 

\subsection{Fragment velocity of the prompt breakup events}
\label{F-prmt}

\begin{figure}[!h] 
\centering
\includegraphics[width=7truecm]{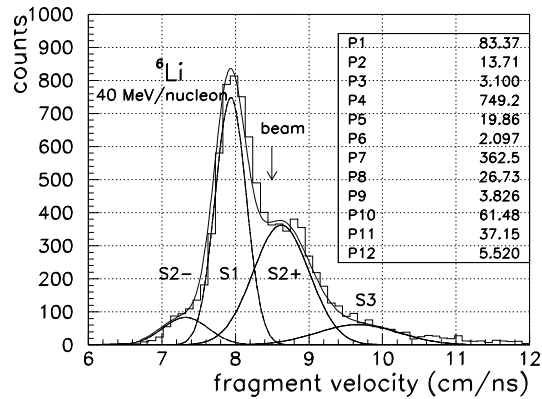}
\caption{Fragment velocity spectrum of the $^6$Li channel at 40~MeV/nucleon energy.}
\label{vFLi}
\end{figure}

In the Coulomb breakup the neutron and the fragment share the decay energy. The fragment
gets the recoil velocity from the neutron emitted. The recoil velocity has a discrete 
value if the neutron is from a resonance decay. Additionally the fragment may get a continuous 
velocity component, the post-breakup acceleration  if the decay happened in the 
Coulomb field of the target nucleus. The post-acceleration depends on the place of the 
decay, on the distance from the target nucleus. The velocity spectrum of the 40~MeV/nucleon 
$^6$Li fragments is shown in fig~\ref{vFLi}. The fragment velocity spectrum  
has a narrow central peak at $v_0$ source velocity and some broader peak pairs
fitted by four Gaussian functions. The S1 narrow peak is the  resonance decay and the 
side peaks are from direct breakup events. The side peaks are asymmetric,  cut by the 
detector acceptances and the magnetic field. 

\begin{figure}[ht] 
\centering
\includegraphics[width=14truecm]{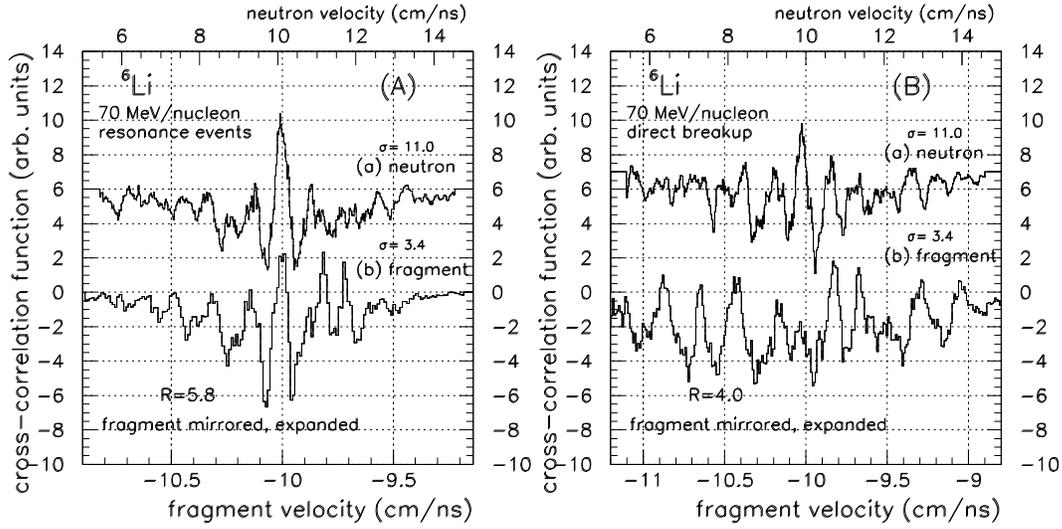}
\caption{Comparison of the cross-correlation function of the $^6$Li fragment  and the neutron velocities of the  resonance (A) and direct breakup (B) events  at 70~MeV/nucleon. The form function is a Gaussian shape function. The fragment spectra are mirrored and expanded to fit the neutron ones. The expansion factors are $R=5.8$ and 4.}
\label{cvnF}
\end{figure}

The cross-correlation analysis of the fragment velocity spectra is shown in fig.~\ref{cvnF}.
The fragment velocity spectra are compared to the neutron ones of the 70~MeV/nucleon $^6$Li+n
resonance (A) and direct breakup (B) events. The fragment velocity spectra (b) are 
mirrored around the $v_0$ source velocity and expanded to fit to the neutron ones (a). 
The expansion factor is $R=5.8$ and 4 of the resonance and direct breakup 
events, respectively. The agreement with the  ratio of the mass numbers ($^6$Li 
and neutron), the momentum conservation verifies that the peaks are true ones, 
the neutron and the fragment are from one decay event, from  a  definite excited state 
of the $^7$Li in both resonance and direct breakup processes. In case of the direct
breakup events the  expansion factor is smaller. Probably the angular range is larger
and the simplified estimation of the track length distorts the velocity scale. 
The comparison of the velocity spectra give similar result at 40~MeV/nucleon and 
for the $^7$Li and $^4$He fragments.

\subsection{Fragment--neutron velocity correlation}
\label{n-F-cor}

\begin{figure}[h] 
\centering
\includegraphics[width=14truecm]{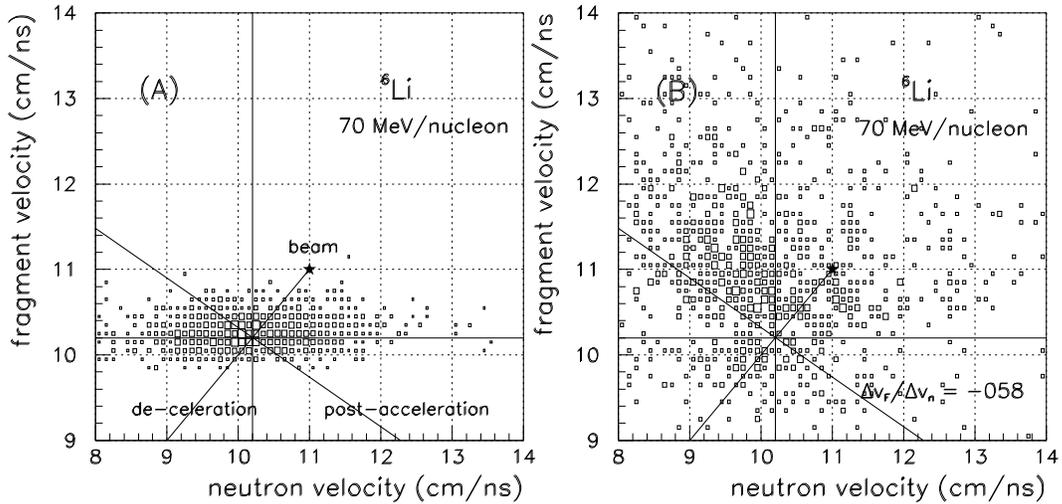}
\caption{Velocity surface  $v_F ~vs~ v_n$ of 70~MeV/nuclon $^6$Li channel. (A) resonance and (B) direct breakup events (in logarithmic scale). The lines are the deceleration,  post-acceleration, and $v_0$ the excitation energy loss corrected beam velocity. The asterisks are the beam velocity.} 
\label{vstxt}
\end{figure}

The velocity correlation of a fragment--neutron pair can be investigated by the
$v_F~{\rm vs}~v_n$ velocity surface plot. It verifies that the decay products 
originated from a common decay process and holds information about the time and place   
dependence of the decay. In fig.~\ref{vstxt} there is shown the scatter plot of the 70~MeV/nucleon
$^6$Li coincidence events.  It is remarkable that there are fragments and neutrons  
faster than the beam. The velocity surface plot of the resonance decay events (A) 
is an isolated peak close to the beam velocity with  $\Delta v_F=\Delta v_n=-0.8$~cm/ns 
velocity deficit. The fig.~\ref{vstxt}(B)  is the velocity surface of the $^6$Li
direct breakup events. The plots are drawn in logarithmic scale to emphasize the structure
of the velocity surface.  The prompt decayed events have a texture, isles. The 
fluctuation suggests that the direct breakup events are also from discrete transitions, 
decayed inside the Coulomb field.

In order to understand the velocity surface plot one has to classify the decay modes.   
As a working hypotheses let us suppose, that the Coulomb dissociation is a two step process. 
The inverse population ratio suggests that in the approaching phase the projectile nucleus 
with valence neutron  cannot decay, but
suffers forced oscillation and surpassing the closest approach point is trapped into
a high energy primary excited state in the continuum: step I. The lifetime of the 
primary state can be large or small compared to the transit time of the projectile
through the Coulomb field of the target nucleus. The primary excited state can decay
prompt or delayed: groups P and D. The secondary  excited state (step II.) may have also 
short or long lifetime. Four variations can be  DI\&DII,  DI\&PII, PI\&DII, and  PI\&PII.
The DI\&DII and the DI\&PII groups are resonance decay events. The fragment is fully
re-accelerated and gets the recoil velocity of the primary and secondary (final) decay. 
In the PI\&DII group the primary excited state decays prompt in the Coulomb field of the 
target nucleus, the primary fragment escapes and decays in-flight. The detected fragment is 
post-accelerated with the Z/A of the primary fragment. It has the primary velocity deficit, 
the velocity excess got in the  post-acceleration, and the recoil velocity of the primary 
and final decay. In the PI\&PII group both decays happened in the Coulomb field simultaneously.
The final fragment (of the secondary decay) is post-accelerated, got recoil velocity and has 
the velocity deficit. 

The neutron velocity will be  the momentary traveling velocity of the projectile, {\it i.e.}
the source velocity, plus the decay velocity decreased by the fragment recoil.  The events 
of the DI\&any group give a spot on the deceleration line at $\Delta v_F=\Delta v_n=v_x$ 
velocity deficit (fig.~\ref{vstxt})(A).The PI\&DII and the PI\&PII 
events decay in the Coulomb field of the target nucleus. If the decay velocity is 
a discrete value, from the decay of a resonance state, the loci of the fragment--neutron 
velocities  will be on a line parallel to the post-acceleration line.  The 
large decay energy primary neutrons shift the post-acceleration line by the recoil 
corrected decay velocity    $v_D'=A_F/(A_F+1)\cdot v_D$ 
to the right and down by the fragment recoil
$-v_D'/A_F$. The above systematics results in a wedge shape velocity 
surface plot.  The slope of the upper edge of the wedge gives the slope of the
post-acceleration line, and the peak of the wedge gives  the decay velocity (fig.\ref{vstxt}(B)).

 The Coulomb energy slows down the projectile and deforms the orbit of the valence neutron. 
The velocity loss is regained, while the deformation energy is lost, frozen into excitation 
energy. Probably the selected impact parameter was  less than 20~fm, the  Coulomb potential of 
the $^8$Li is larger,  for the velocity deficit itself is $v_x=-0.8$~cm/ns which corresponds to 
21~MeV excitation energy.

\begin{figure}[ht] 
\centering
\includegraphics[width=10truecm]{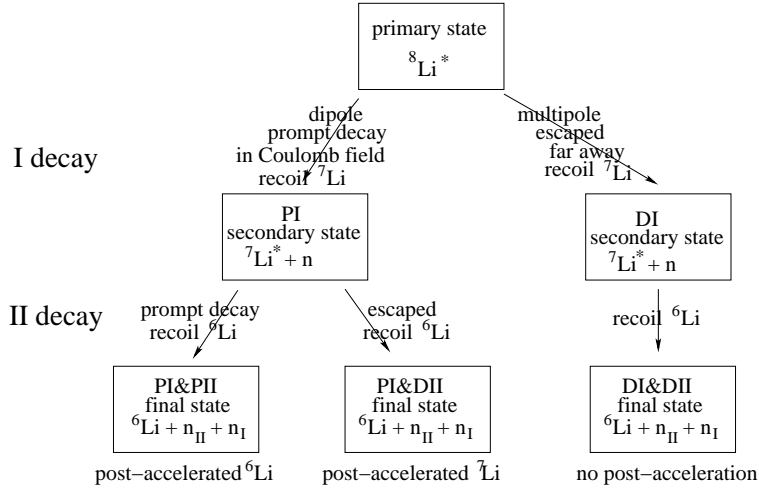}
\caption{Flow diagram of the $^6$Li+2n Coulomb  dissociation. The decay modes are: DI\&DII double delayed, resonance decay, PI\&PII double prompt decay,  PI\&DII  primary prompt, secondary delayed, post-accelerated resonance  decay.}
\label{vsLi6m}
\end{figure}

From the observations above one can conclude that the secondary decay inherits the lifetime 
of the primary excited state. The same final transitions may happen prompt or delayed. 
That results in the resonance decay and direct breakup  processes.
The decay mode is determined by the multipolarity of the orbit of the 
valence neutron. Probably the dominantly dipole excitations decay prompt and the 
multipole orbits live long. The decay modes are summarized in fig.~\ref{vsLi6m}.

Knowing the mass number of the fragment group, from the slope of the post-acceleration line 
one can get the Z charge of the fragment. The energy loss or gain of an ion in the Coulomb field at 
a distance $r$ from the target nucleus is proportional with the Z/A ratio of the ion.
If in the decay the Z/A ratio changes the fragment gets net velocity change. The 
measured quantities are the $v_n$ neutron and $v_F$ fragment velocities. The velocity 
change is the difference from the source velocity. The net velocity excess is the
post-breakup acceleration. Supposing that in the moment of the decay the neutron velocity 
was equal with the projectile velocity, {\it i.e.} $\Delta v_n(r)=\Delta v_p(r)$, 
the neutrons preserve the velocity in the moment of the decay. Then  the  velocity excess, 
the post-breakup acceleration of the fragment is
$\Delta v_{pa}=\Delta v_F-\Delta v_p=\Delta v_F-\Delta v_n$, and the relative 
velocity change of the fragment and the neutron is the slope of the post-breakup acceleration 
line $m=\Delta v_{pa}/\Delta v_n=\Delta v_F/\Delta v_n-1$. On the other hand it was shown 
that the relative velocity increment  can be given with the Z/A ratio of the projectile and 
the fragment: $m=-\sqrt{\varepsilon }=-\sqrt{Z_F/Z_p\cdot A_p/A_F-1}$ (Sec.~7). 

\subsubsection{$^6$Li+2n channel}
\label{vs-Li6}

\begin{figure}[h] 
\centering
\includegraphics[width=14truecm]{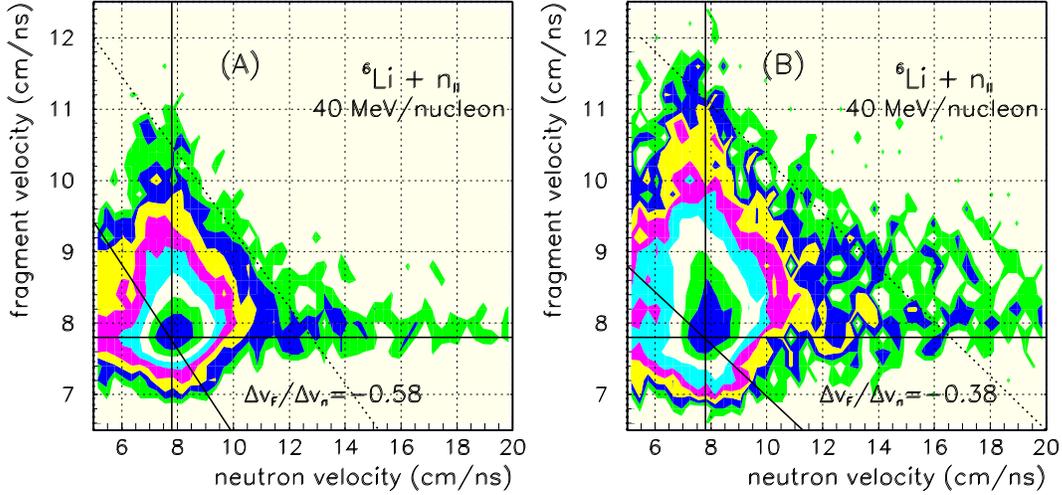}
\caption{Contour plots of the velocity surface  $v_F ~vs~ v_n$ of 40~MeV/nucleon $^6$Li fragments in logarithmic scale. (A) secondary, low velocity neutrons (single counts suppressed), (B) primary neutrons.  The lines are the excitation energy loss corrected  beam velocity and  the post-breakup acceleration lines. The dotted lines show the upper edge of the wedge.}
\label{vsL6}
\end{figure}

The velocity surface of the 40~MeV/nucleon $^6$Li+n coincidence events is shown in 
fig.~\ref{vsL6} . The plot (A) is a contour plot with single counts suppressed,  
to show the  structure of the velocity surface of the $n_{II}$ low velocity secondary neutrons. 
It has a wedge shape. The lines are the post-acceleration and the cross the beam velocity
corrected by the excitation velocity loss. 

In the velocity surface there are fragments and neutrons much faster than the beam. 
The projectile is slowed down in the approaching  phase,  and surpassing the closest 
approach point will be re-accelerated.
The fragments of the prompt decay events get significant post-breakup acceleration.
From the fig.~\ref{vsL6}(A) the slope of the upper edge of the wedge $m=-0.58$ agrees
with the post-acceleration of the  $^6$Li ($m=-\sqrt{8/6-1}=-0.58$). That means
the $^8$Li projectile decayed somewhere in the Coulomb field and 
the second decay followed the primary decay in no time and the $^6$Li was accelerated 
(PI\&PII). The plot (B) contains the high velocity primary neutrons too. The slope 
of the upper edge  of the wedge is $m=-0.38$, which refers to the post acceleration of the
intermediate $^7$Li$^*$ fragment $m=-\sqrt{8/7-1}=-0.38$.  The primary decay was prompt 
and the secondary one was delayed (PI\&DII). The detected $^6$Li fragment was accelerated
as $^7$Li. The upper edge of the wedge is a broad line formation, one can estimate 
the middle, the ridge  of the line. In the plot (B)  the peak of the wedge  of the 
primary neutrons is about  $v'_D=6$--7~cm/ns, which corresponds to $E_D=22$~MeV 
decay energy.

\begin{figure}[!h] 
\centering
\includegraphics[width=14truecm]{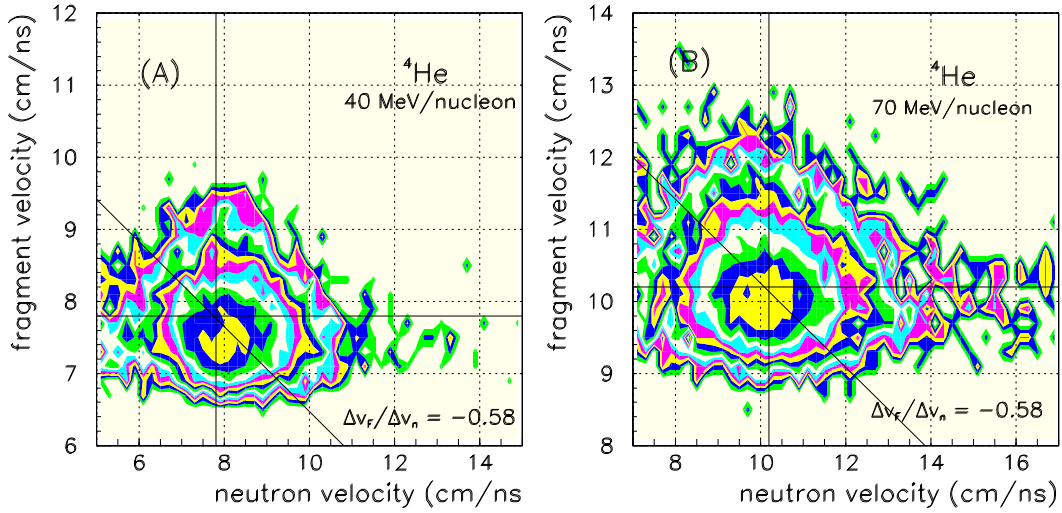}
\caption{Velocity surface plots  $v_F ~vs~ v_n$ of $^4$He+t+n channel  (A) 40 and  (B) at 70~MeV/nucleon.}
\label{vsHe4}
\end{figure}

The $^7$Li channel has no high velocity events above the beam velocity.  
Probably the projectile lost the excitation energy by $\gamma $-ray cascade, 
missed the channels opening and decayed from a low lying  unbound resonance 
state of the $^8$Li.  At both energies  
the $^7$Li is the end station of the relaxation of the $^8$Li.
In fig.~\ref{vsHe4} the $^4$He  velocity surface plots are compared at the two energies. 
The slope of the post-acceleration lines is -0.58. It is equal with  
$\Delta v_F/\Delta v_n=-\sqrt{2/4\cdot 8/3 -1}=-0.58$, the post-acceleration of 
the $^4$He. 

The velocity surface analysis  supports the identification  of the 
$^{6,7}$Li and $^4$He fragments.
The $^2$H and $^6$He channels are discussed in detail for they serve also valuable
arguments  for the particle identification and the reaction mechanism.

\subsubsection{$^2$H+$^5$He+n channel}
\label{h2}

\begin{figure}[h] 
\centering
\includegraphics[width=7truecm]{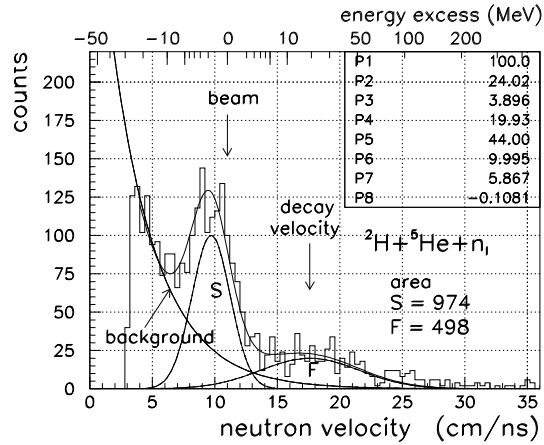}
\caption{Neutron velocity spectrum of the $^2$H+$^5$He+n$_I$ reaction channel at 70~MeV/nucleon.}
\label{vnd}
\end{figure}

The $^2$H--n coincidence  events are the direct witnesses of the primary high excitation 
of the projectile. The threshold of the reaction  is  11.65~MeV. 
The neutron velocity spectrum of the $^2$H channel is shown in fig.~\ref{vnd}.
The spectrum is unfolded by two Gaussian functions on an exponential background.
The velocity  increment of the high energy neutrons   is about $6.2\pm4$~cm/ns, 
which corresponds  to 20.5~MeV decay energy. 


\begin{figure}[h] 
\centering
\includegraphics[width=8truecm]{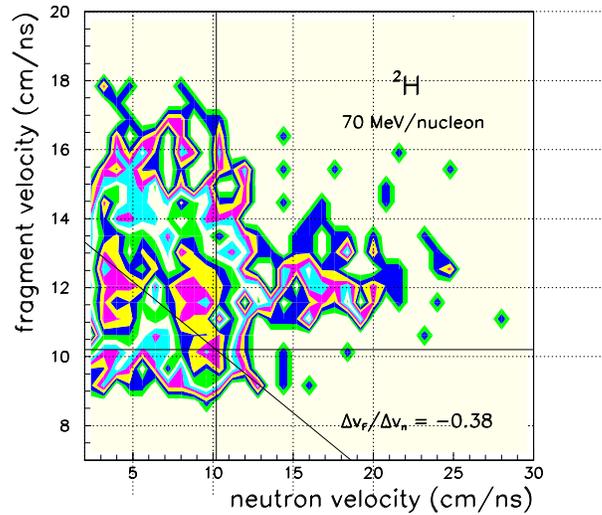}
\caption{Velocity surface  of  $^2$H+$^5$He+n$_I$ reaction channel at 70~MeV/nucleon.}
\label{vsd}
\end{figure}

The velocity surface of the $^2$H--n coincidence events is shown in fig.~\ref{vsd}. 
The slope of the upper edge is $m=-0.38$. This support that the reaction is a two step
decay: $^8$Li$\rightarrow ^7$Li+n$_I  \rightarrow ^5$He+$^2$H+n$_I$, and the  
$^7$Li intermediate fragment is accelerated: $\Delta v_F/\Delta v_n=-\sqrt{8/7-1}=-0.38$. 
The fragment velocity excess goes up to 6~cm/ns.  Probably the $^2$H channel is  
PI\&DII type prompt decay, close to the closest approach point.  The measured velocity 
excess of the $^2$H is the sum of the post-acceleration of the $^7$Li primary fragment 
and the recoil velocity in the secondary $^7$Li$\rightarrow ^5$He+$^2$H decay. 


\subsubsection{$^6$He channel}
\label{he6}
The large yield of the $^6$He channel is a strong argument of the excitation into 
the giant resonance region. The threshold energy is 9.78 and  13.70~MeV. The $^6$He channel is 
the most intensive reaction channel at 40~MeV/nucleon, admitting that those are 
over represented accidental coincidences.              

\begin{figure}[h] 
\centering
\includegraphics[width=7truecm]{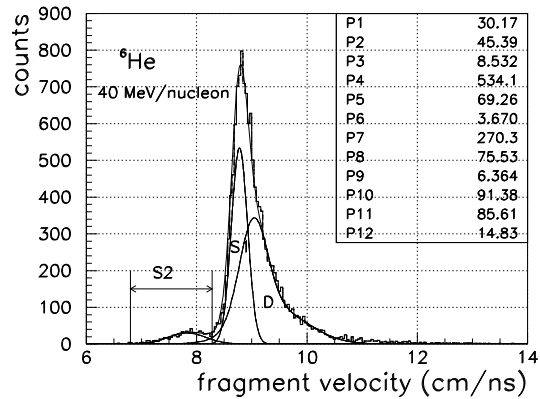}
\caption{$^6$He velocity spectrum at 40~MeV/nucleon. The components are Gaussian fit results: S1 
resonance decay, D direct breakup. (Fit parameters in 0.03~cm/ns units.) The S2 side peak is the $^7$Li$\rightarrow \beta +^7$He$\rightarrow ^6$He+n reaction.}
\label{vFHe6}
\end{figure}

The velocity spectrum of the  $^6$He  (fig.~\ref{vFHe6}) is an asymmetric peak.
It is unfolded by 4 Gaussian functions. There is a narrow  peak S1 at 8.8~cm/n  with  
FWHM=0.35~cm/ns. It refers to a long lived  resonance decay, although it is strongly
cut by the CRDC telescope. The broader high energy tail D (the sum of 2 Gaussian 
functions) belongs to the post-accelerated, prompt decay in the Coulomb 
field of the target nucleus. And there is  a small side peak S2 at about 7.8~cm/ns. 

\begin{figure}[h] 
\centering
\includegraphics[width=14truecm]{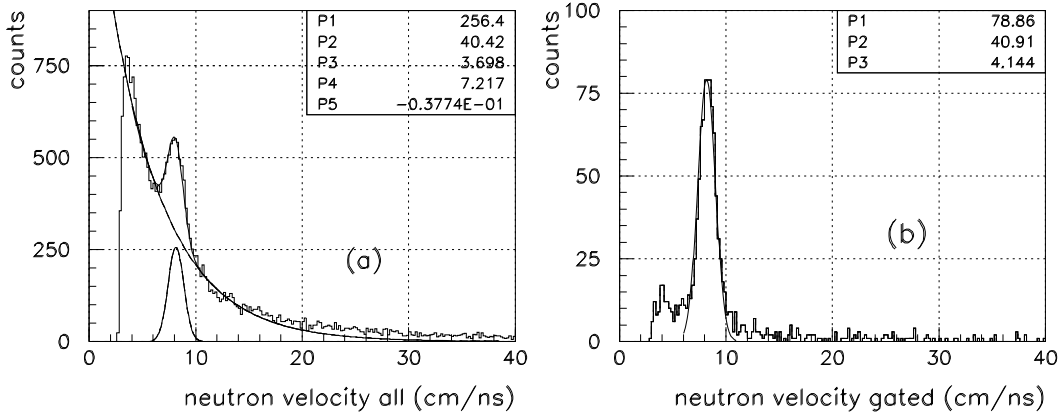}
\caption{MoNA velocity spectra. (a) in coincidence with $^6$He, (b) neutron velocity spectrum cut 
by $^7$He gate  (S2 in fig.~\ref{vFHe6}).Fit parameters in 0.2~cm units.}
\label{vnHe6}
\end{figure}

The neutron velocity spectrum in coincidence with the $^6$He fragments is an exponential 
decreasing one with a small bump at 8.1~cm/ns (fig.~\ref{vnHe6}(a)). The exponential spectrum 
is an accidental coincidence spectrum because  of the missing MoNA fast clear.  
The $^6$He has no neutron partner, probably it is from the  $^8$Li$\rightarrow ^6$He+d 
reaction channel. The small bump of the neutron velocity spectrum is in coincidence with 
the S2 side peak of the $^6$He velocity spectrum. 

\begin{figure}[ht] 
\centering
\includegraphics[width=14truecm]{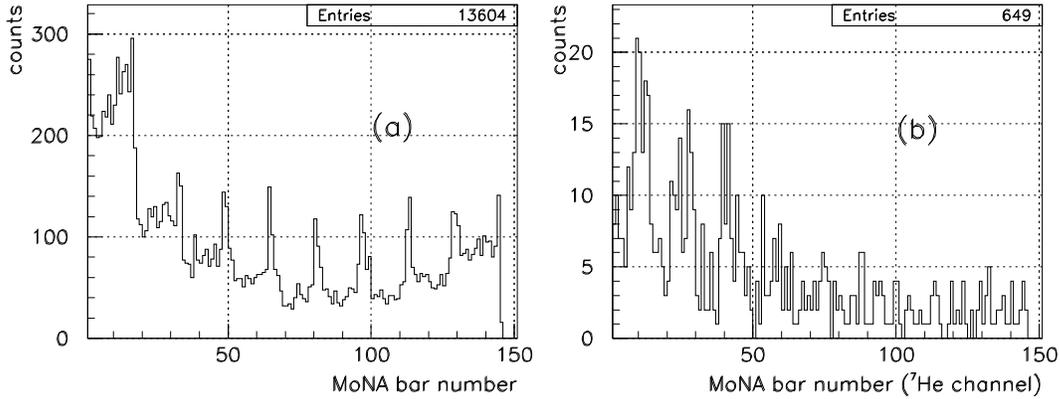}
\caption{MoNA bar frequency in coincidence with $^6$He fragments at 40~MeV/nucleon.  (a) all coincidence events, (b) gated by the S2 condition.}
\label{BA6}
\end{figure}

The MoNA bar frequency spectrum verifies that the neutrons are accidental ones (fig.~\ref{BA6}(a)). 
The bar distribution shows a decreasing 16-th periodic structure, but the peaks are flat with a sharp
peak in the lowest bars (16$^{th}$, 32$^{nd}$, {\it e.t.c}) scattered in from the floor
and from the wall behind the  detector. 
In the $^6$He--neutron coincidences the neutrons belong to a foreign fragment, to an other 
reaction in the time interval of the coincidence gate. 

The cyclotron period was 45.82~ns (21.8259 MHz).
The TDC range was 320~ns, about 7 periods. The coincidence gate covered  3 cyclotron periods. 
The beam intensity was 150000/s, the average ion rate is 6.6~ion/$\mu$s, which gives
144~cycle/ion frequency. So at the beam intensity the cycles generally  were empty, even 
probably there were a few percent spills with multiple ions in it. 
Supposedly the extraction from the ion source determines the beam structure. 
Really every event is fragment--MoNA coincidence event and the true coincident
neutrons can be identified with valid TOF value, while the accidental neutrons generally went 
to overflow. The bar multiplicity of the true coincidence neutrons is 1--3 
(fig.~\ref{MU}(b)). 

The small bump at 7.8~cm/ns in the $^6$He velocity spectrum  (fig.~\ref{vFHe6}) 
belongs to true fragment--neutron  coincidences. The neutron velocity spectrum 
gated by the $^6$He fragments signed by S2 
is shown in fig.~\ref{vnHe6}(b). It is a narrow neutron peak of FWHM=1~cm/ns. They are 
neutrons from the target in the center of the walls (fig.~\ref{BA6}(b)).  
The MoNA hits can be neutrons --- at these energies --- from the reaction
$^8$Li$^*\rightarrow ^7$Li+n$_I\rightarrow^7$He+$\beta ^+$+n$_I\rightarrow^6$He+n$_{II}+\beta ^+$
+n$_I$.
The decay energy of the $^7$He$\rightarrow ^6$He+n$_{II}$ reaction is $E_D=0.44$~MeV.
This decay channel is  added to  the decay scheme of the $^8$Li shown in fig.~\ref{Li8}.
The $^6$He+d channel and the $^6$He+n channel  both are resonance decay events with 
narrow fragment velocity spectrum.

\begin{figure}[ht] 
\centering
\includegraphics[width=7.5truecm]{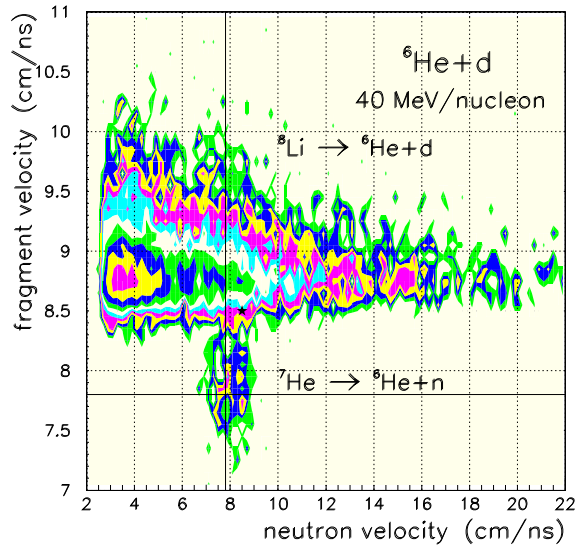}
\caption{Velocity surface of the $^6$He+MoNA coincidences.}
\label{vsHe6}
\end{figure}

The $v_F ~vs~v_n$ velocity surface is shown in fig.~\ref{vsHe6} of the $^6$He ions.
The $^6$He velocity of the accidental coincidence events have a peak at 8.8~cm/ns,  
while the true coincident $^6$He-s are at $v_0=7.8$~cm/ns.  The velocity spectrum  
of the $^6$He of the deuteron channel is shifted up because of the recoil of
the  $^6$He by the 1.04~MeV  deuteron. Further the acceptance of the FPD selects
the fragments recoiled into the forward direction. The velocity surface  plot 
verifies the fragment identification and the presence  of the very intensive 
$^6$He fragments  supports the leakage decay model of the  Coulomb disintegration. 


\section{Leakage model of the decay and localized valence neutron model}
\label{lkg}

Accepting the fragment identification, the high threshold and  the inverse  
population ratio of the  $^7$Li+n, $^6$Li+2n, $^4$He+t+n, $^6$He+2n+$\beta ^+$, $^6$He+d,
and $^2$H+n+$^5$He  reaction channels, the equal velocity deficit,  the identical 
structure of the  velocity spectra of the prompt and delayed reaction products
suggest that  the Coulomb dissociation of a neutron rich light nucleus is a two 
step process. In the approaching phase the projectile is braked down by the 
Coulomb field of the target nucleus, and  the orbit of the valence neutron is
strongly deformed. The valence neutron gets forced oscillation. In the increasing 
Coulomb field  the $^8$Li  is not  in its eigenstate 
and cannot decay. Similarly to the instantaneous model \cite{BaBe}. 
Passing the turning point, released, the ion will be trapped into 
a highly excited state which fits to the oscillation mode.  The excited projectile 
traveling further will be re-accelerated and decays with the lifetime of the resonance 
state, emits a neutron or $\gamma $-rays. 
The dominantly single-particle, dipole excited states decay prompt,
in the Coulomb field of the target nucleus, while the multipole excited states 
live long  until the energy is concentrated to the valence neutron
and decay departing, in flight  emitting a mono-energetic neutron. 
The above reaction mechanism  can be called the {\it leakage model 
of the decay}. The highly excited primary states decay by high energy neutron emission 
or are degraded by $\gamma $-ray cascade. The rest of the beam will feed the lower 
energy channels. This results in the {\it inverse population ratio}, and that the share 
of the simplest  $^8$Li$^* \rightarrow ^7$Li+n  channel is only 14 and 16\% at 40 
and 70~MeV/nucleon.
  
The collision of intermediate and high energy projectile is a sudden process.
The orbital period of the valence nucleon is larger than the transit time of the
projectile through the Coulomb field of the target nucleus.  The interaction radius of 
the $^8$Li is $r=2.36\pm0.02$~fm \cite{AS2}. Supposing that the mean radius does not 
changes the $E_x=18$~MeV excitation energy corresponds to the frequency increment 
of the orbital motion.
$$
E_x=\hbar \omega~, \hskip 0.3truecm \nu =4.47\cdot 10^{21}~{\rm s^{-1}, ~~  and } \hskip 0.3truecm \Delta v_n=6.6~{\rm cm/ns}
$$ 
in agreement with the measured velocity increment (peak F in fig.~\ref{vnd}). 
It can be supposed that during the 
impact the valence neutron stays in the forward or backward hemisphere of the projectile.
According to the {\it localized valence neutron} assumption in the forward hemisphere 
the projectile gets a prolate deformation, dominantly dipole and  single-particle  excitation,
and in the backward hemisphere oblate deformation, multipole  and  collective excitation.
Basically two kinds of primary  states can be excited  corresponding to the
position of the valence neutron during the impact: a short and a long living
excited state according to the dipole or multipole deformation. The different 
reaction channels open through one of the primary excited states. 
The secondary breakup generally inherits the lifetime of the primary
excited state. Therefore the fragments have a delayed narrow resonance like group 
in the velocity spectrum 
and a prompt broader one extended by the post-breakup acceleration.  The multipolarity 
of the primary excited state is determined by the multipolarity of the orbit.

According to the Heisenberg Uncertainty Principle the momentum uncertainty of the
localized neutron ($\Delta x=2.4$~fm) is $\Delta p=84$~MeV/c. The experiment  could 
be close to this  limit  for at 40~MeV/nucleon the separation of the two mechanisms 
is less sharp than at 70~MeV/nucleon. At the impact parameter scale the wedge shape 
of the velocity surface, the post-acceleration gives information about the position or  
the time of the decay  in the Coulomb field of the target nucleus. The post-acceleration 
is a clock of the reaction \cite{BeBa}.

\section{Summary}
\label{sum}

The paper presents the analysis of experimental data from  Coulomb dissociation of 
$^8$Li on Pb target  at 40 and 70~MeV/nucleon energy.  The $^{6,7}$Li, $^{4,6}$He, 
and $^2$H  fragments are separated.

The two mechanisms of the Coulomb dissociation, the soft dipole resonance and
the direct breakup are demonstrated.  The resonance decay, discrete low decay energy 
neutrons are identified in the $^8$Li$\rightarrow ^6$Li+2n dissociation and the
post-breakup acceleration of the prompt, direct breakup  fragments is verified.
It was found that the structure of the neutron velocity spectra of the two 
reaction mechanisms is identical.  It  has to be  supposed that he Coulomb dissociation 
is a two step process. The primary excited states may have  short or long lifetime, 
so the neutrons of the secondary decay, from the same transitions, can appear 
prompt or delayed related to the projectile. The secondary decay inherits the lifetime of 
the primary decay. The decay, primary or secondary, in the Coulomb  field of the target
nucleus  results in  the post-breakup acceleration of the fragment.

The threshold of the reaction channels is high, more than 10~MeV. The velocity loss
of the resonance decay products corresponds to about 21~MeV excitation energy at 
the given magnetic field and impact parameter set, at both energies and for every 
reaction channels. Probably the primary  excited states are common for the channels. 
The secondary excited states are fed from up downwards  by the rest of the beam. 
It is called {\it leakage decay model} of the Coulomb dissociation. 
That results in the {\it inverse population ratio} of the reaction channels, {\it i.e.}  
the higher  the energy of the secondary excited state, the larger is the 
probability of the decay through that channel.

The intermediate and high energy collisions are sudden reactions, the orbital period 
of the valence neutron is  larger than the transit time of the projectile through
the Coulomb field of the target nucleus.  The sudden reaction allows the {\it localized
valence neutron model}, that the valence neutron stays in the forward or backward hemisphere 
of the projectile during the impact. Corresponding to the position of the valence neutron
the deformation of the forced oscillation will be  prolate, dominantly dipole or oblate, 
multipole  oscillation, single-particle or collective excitation. which determines 
the lifetime of the excited state.

The conclusion of the paper is: the data give experimental evidence that a quantum system 
under the effect of an increasing  long range external force gets forced oscillation 
and cannot fall into an eigenstate of its subsystem, but the force pushes it through several 
states up to the giant resonance region. In the Pb+core+valence neutron+electromagnetic field quantum
system the orbit of the valence neutron deforms quasi continuously and passing the closest approach 
point, released the nucleus is trapped into a highly excited state which fits to the quantum 
numbers of the oscillation. The highly excited state decays with own lifetime 
delayed or prompt giving the soft dipole resonance and the so called 'direct breakup' decay 
mechanisms of the Coulomb dissociation. The direct breakup is really  also  resonance decay 
of the low lying   unbound resonance state but prompt even in the strong Coulomb field.

\ack

The author is indebted to Vladimir Zelevinsky for his stimulation 
and valuable discussions, to Aaron Galonsky for his substantial 
remarks and advices,
and to Rudolf Izs\'ak for his cooperation  and technical help. 
The author is grateful to the MoNA collaboration 
that made available  the experimental data.
The participants of the MoNA collaboration are:
M. Csan\'ad$^1$, F. De\'ak$^1$, \'A. Horv\'ath$^1$, R. Izs\'ak$^1$,
\'A. Kiss$^1$, G. I. Veres$^1$, 
Z. Seres$^2$,
T. Baumann$^3$, D. Bazin$^3$, N. Frank$^3$, A. Gade$^3$,
D. Galaviz$^3$, A. Galonsky$^3$, A. Schiller$^3$, M. Toennessen$^3$,
P. DeYoung$^4$, 
W. A. Peters$^5$,
K. Ieki$^6$, R. Sugo$^6$, 
T. Fukuchi$^7$,
Zs. F\"ul\"op$^8$, 
C. A. Bertula\-ni$^9$, 
H. Schelin$^{10}$,
N. Carlin$^{11}$, and 
C. Bordeanu$^{12}$. 
The support of the Hungarian OTKA grant No. T049837 and
the National Science Foundation under Nos. Phy01-10253, PHY03-54920, 
and PHY04-56463 are gratefully acknowledged. \hfill\break

{\small 
\it 
\noindent
$^1${Department of Atomic Physics, E\"otv\"os 
Lor\'and University, H-1117 Budapest, Hungary}\hfill\break
\noindent
$^2${Institute for Particle and Nuclear Physics, Wigner Research Centre 
for Physics, H-1525 Budapest, Hungary}\hfill\break
\noindent
$^3${National Superconducting Cyclotron Laboratory, Michigan State 
University, East Lansing, Michigan 48824, USA}\hfill\break
\noindent
$^4${Department of Physics and Egineering, Hope College, Holland,
Michigan 49423, USA}\hfill\break
\noindent
$^5${University of Tennessee, ORNL, Physics Division, Oak Ridge, TN 37831, USA} \hfill\break
\noindent
$^6${Department of Physics, Rikkyo University, 3 Nishi-Ikebukuro, Toshima,
Tokyo 171, Japan}\hfill\break
\noindent
$^7${RIKEN Center for Life Science  Tchnologies, Kobe, Hyogo 650-0047, Japan}\hfill\break
\noindent
$^8${ATOMKI Institute for Nuclear Research, H-4001 Debrecen, Hungary}
\hfill\break
\noindent
$^9${Department of Physics and Astronomy, Texas A\&M 
Univer\-sity--Commerce, Commerce, Texas 75429, USA}\hfill\break
\noindent
$^{10}${Federal University of Technology--Parana, 80230-901 Curiti\-ba, 
Paran\'a, Brail}\hfill\break
\noindent
$^{11}${Instituto de F\'{\i}zica, Universidade de S\~ao Paulo, 05315-970 S\~ao 
Paulo, Brazil}\hfill\break
\noindent
$^{12}${National Institute for Physics and Nuclear Engineering - Horia Hulubei,
M\u{a}gurele-Bucharest, Romania, 07125}
}

\vskip 2pc
\noindent
\bf{References}

\end{document}